\documentclass[preprint2]{aastex}
\usepackage{times}
\usepackage{color}

\begin{document}

\title{Far-ultraviolet fluorescent molecular hydrogen emission map of the Milky Way Galaxy}

\author{Young-Soo Jo\altaffilmark{1}, Kwang-Il Seon\altaffilmark{2,3}, Kyoung-Wook Min\altaffilmark{1}, Jerry Edelstein\altaffilmark{4}, Wonyong Han\altaffilmark{2}}
\email{email: stspeak@gmail.com}
\altaffiltext{1}{Korea Advanced Institute of Science and Technology (KAIST), 291 Daehak-ro, Yuseong-gu, Daejeon, Korea 305-701, Republic of Korea}
\altaffiltext{2}{Korea Astronomy and Space Science Institute (KASI), 776 Daedeokdae-ro, Yuseong-gu, Daejeon, Korea 305-348, Republic of Korea}
\altaffiltext{3}{Astronomy and Space Science Major, Korea University of Science and Technology, 217 Gajeong-ro, Yuseong-gu, Daejeon, Korea 305-350, Republic of Korea}
\altaffiltext{4}{Space Sciences Laboratory, University of California, Berkeley, CA, USA.}

\begin{abstract}
We present the far-ultraviolet (FUV) fluorescent molecular hydrogen (H$_2$) emission map of the Milky Way Galaxy obtained with \textit{FIMS/SPEAR} covering $\sim$76$\%$ of the sky. The extinction-corrected intensity of the fluorescent H$_2$ emission has a strong linear correlation with the well-known tracers of the cold interstellar medium (ISM), including color excess \textit{E(B\textendash V)}, neutral hydrogen column density \textit{N}(\mbox{H\,{\sc i}}), and H$\alpha$ emission. The all-sky H$_2$ column density map was also obtained using a simple photodissociation region model and interstellar radiation fields derived from UV star catalogs. We estimated the fraction of H$_2$ ($f_{H2}$) and the gas-to-dust ratio (GDR) of the diffuse ISM. The $f_{H2}$ gradually increases from $<$1$\%$ at optically thin regions where \textit{E(B\textendash V)} $<$ 0.1 to $\sim$50$\%$ for \textit{E(B\textendash V)} = 3. The estimated GDR is $\sim$5.1 $\times$ 10$^{21}$ atoms cm$^{-2}$ mag$^{-1}$, in agreement with the standard value of 5.8 $\times$ 10$^{21}$ atoms cm$^{-2}$ mag$^{-1}$.
\end{abstract}


\section{Introduction}

Hydrogen is the most abundant element in the Milky Way Galaxy. It accounts for $\sim$90$\%$ of the interstellar medium (ISM) by number or $\sim$70$\%$ by mass \citep{bin87}. About 20$\%$ of interstellar hydrogen is in the form of molecular hydrogen (H$_2$), which is an important tracer of star-forming regions or molecular clouds \citep{gou63,hol71,nak03,nak06,shu82}. However, it is difficult to observe H$_2$ directly because of the absence of a permanent electric dipole moment. In most cases, the amount of H$_2$ is inferred from the amount of dust present using the gas-to-dust ratio (GDR) or from radio observations of carbon monoxide (CO) using the CO-to-H$_2$ conversion factor \textit{X}$_{CO}$. Although CO is much less abundant than H$_2$ ($\sim$0.02$\%$ of H$_2$), it can be easily observed via the $^{12}$C$^{18}$O \textit{J} = 1$\rightarrow$0 emission line at 2.6 mm. Therefore, \textit{X}$_{CO}$, defined as the ratio of the H$_2$ column density (\textit{N}(H$_2$)) to the integrated CO intensity (\textit{W}$_{CO}$), has been used to derive the amount of H$_2$ in molecular clouds \citep[e.g.,][]{nak06}. The \textit{X}$_{CO}$ factor is usually assumed a constant; e.g., \citet{dam01} measured the Galactic mean value of \textit{X}$_{CO}$ as (1.8 $\pm$ 0.3) $\times$ 10$^{20}$ cm$^{-2}$ K$^{-1}$ km$^{-1}$ s. However, in principle, \textit{X}$_{CO}$ can vary depending on the local conditions of the ISM, e.g., the cloud density and excitation temperature. Indeed, many studies have shown that \textit{X}$_{CO}$ can be substantially different from the Galactic mean value and depend on the interstellar environment \citep{dev87,lee14,mag88,pin08,pin10}. The amount of H$_2$ can also be derived using \textit{N}(\mbox{H\,{\sc i}}) and \textit{E(B\textendash V)} under the assumption that the GDR is a constant, i.e., $\langle$\textit{N}(\mbox{H\,{\sc i}} + H$_2$)/\textit{E(B\textendash V)}$\rangle$. The standard value of the GDR for the Milky Way Galaxy is 5.8 $\times$ 10$^{21}$ atoms cm$^{-2}$ mag$^{-1}$ \citep{boh78,rac09}. However, this value was obtained from the observations of a finite number of sight lines. Ideally, direct observations would be the best way to determine the amount of H$_2$ rather than using the GDR or \textit{X}$_{CO}$.

H$_2$ in the ground state absorb far-ultraviolet (FUV) photons of wavelength $\lambda >$ 912 {\AA}, which excite the molecules into electronically excited states. Approximately 10$\%$ of the excited H$_2$ dissociate and the remaining 90$\%$ de-excite to vibrationally excited states of the ground electronic state ($X^1\Sigma^+_g$) and emit FUV emission lines at wavelengths between 912 and 1700 {\AA} via transitions from excited electronic states ($B^1\Sigma^+_u$, $C^1\Pi^u$) to the ground state. The transitions within the vibrational rotational energy levels of the ground electronic state result in quadrupole transition lines at the near- and mid-infrared wavelengths \citep{pak03,ste89}. Ground-based telescopes have directly detected H$_2$ fluorescence emission lines in the near-infrared (NIR) in a small number of sight lines toward external galaxies \citep{mar99,pak04} or molecular clouds in our Galaxy \citep{car08,fra16,zha15}. FUV fluorescent H$_2$ emission lines in the 1350--1750 {\AA} wavelength range have also been observed in a few targets: the Taurus molecular cloud \citep{lee06}, the Ophiuchi cloud complex \citep{lim15}, the Orion Eridanus superbubble \citep{jo11,ryu06}, and the local supershell GSH 006-15+7 \citep{jo15}. FUV fluorescent H$_2$ emission lines at shorter wavelengths (900--1400 {\AA}) have been observed in the emission nebulae IC 405 \citep{fra04} and IC 63 \citep{fra05} and the Orion nebula \citep{fra05m}. However, there have been no NIR or FUV observational studies of fluorescent H$_2$ emission lines over the entire sky.

We present our H$_2$ fluorescence emission map of a substantial fraction of the sky obtained with the Far-ultraviolet Imaging Spectrograph (\textit{FIMS}), also known as Spectroscopy of Plasma Evolution from Astrophysical Radiation (\textit{SPEAR}) \citep{ede06a,ede06b}. Construction of the map is discussed in Section 2. Section 3 presents the all-sky \textit{N}(H$_2$) map constructed by applying the photodissociation region (PDR) model to the extinction-corrected H$_2$ fluorescence emission map. The estimated GDR for the diffuse ISM in the Milky Way Galaxy is also discussed in Section 3. A summary of the study is given in Section 4.


\section{Observations and Construction of the Map}

\subsection{Construction of 3-Dimensional Data Cube}

\textit{FIMS/SPEAR} was the main payload on the first Korean scientific satellite, \textit{STSAT-1}, that was launched into a $\sim$700-km Sun-synchronous orbit on 2003 September 27. \textit{FIMS/SPEAR} was specifically designed to provide a spectral imaging survey of diffuse FUV radiation from the ISM. The \textit{FIMS/SPEAR} data demonstrated that diffuse FUV emission lines emanate from various atoms and H$_2$ in a variety of interstellar environments. The instrument has surveyed $\sim$76$\%$ of the sky and has conducted pointed observations toward numerous preselected targets. \textit{FIMS/SPEAR} is a dual-imaging spectrograph composed of two channels, a short-wavelength channel (\textit{S} band; 900--1150 {\AA}, 4\fdg0 $\times$ 4\farcm6 field-of-view) and a long-wavelength channel (\textit{L} band; 1350--1750 {\AA}, 7\fdg4 $\times$ 4\farcm3 field-of-view). Both channels have a spectral resolution of $\lambda/\Delta\lambda \sim$ 550 and an imaging resolution of $\sim$5\arcmin. Each channel is composed of a collecting mirror, a diffraction grating, and a photon-counting microchannel-plate detector. For our map, only the \textit{L}-band data were used because the \textit{S}-band data were badly contaminated by the geocoronal emission lines and low detector sensitivity.

\textit{STSAT-1} orbited around the Earth $\sim$14 times a day, with 7 orbits used for astronomical observations; thus, 1523 orbits were devoted to astronomical observations during the lifetime of the mission (2003 November to 2005 May). After correction of the attitude information, data from only 1244 orbits were of sufficient quality for use in the present analysis. Of these, data from 206 orbits obtained with the 10$\%$ shutter aperture mode were excluded because of a low signal-to-noise ratio (SNR), leaving useful data from only 1038 orbits obtained with the 100$\%$ shutter aperture mode. Of those 1038 orbits, data from 842 were obtained using the ``sky-survey mode,'' which scanned a 7\fdg4 $\times$ 180$\degr$ swath of the sky from the north ecliptic pole to the south ecliptic pole, and the data from the remaining orbits were obtained using the ``targeting mode,'' in which the \textit{FIMS/SPEAR} pointed toward preselected targets such as the Eridanus superbubble, Taurus molecular clouds, the Monogem ring, and Vela supernova remnants. More details about the \textit{FIMS/SPEAR} observations can be found in \citet{seo11}, which contains the all-sky survey map of the FUV continuum background obtained by \textit{FIMS/SPEAR}, excluding data from the orbits with incorrect attitude information. In the present study, we recovered most of the data removed in \citet{seo11} using an elaborate attitude correction procedure. 

With this expanded dataset of 1038 orbits, we constructed a data cube for diffuse emissions only, excluding direct photons that originate from point sources. The removal procedure of bright point sources is as follows. First, we developed a photon count-rate map that was pixelated using the HEALPix scheme \citep{gor05} with a resolution parameter $N_{side}$ = 512 that corresponds to a pixel size of about 6.9 arcmin. We noted that the photons from bright point sources were scattered by the instrument and contaminated the pixels around those corresponding to the point sources. Hence, we performed two-dimensional Gaussian fitting for the bright pixels whose photon count-rates were higher than the median photon count-rates of the sub-region of 100 arcmin $\times$ 50 arcmin around each pixel of the image by a factor of three or more. We also removed all the pixels within the oval regions corresponding to the three-sigma widths derived from the fitting for each bright source. The construction of the final three-dimensional (3D) \textit{L}-band data cube was performed with a resolution parameter $N_{side}$ = 64, resulting in 49,152 spatial pixels in total, which corresponded to a pixel size of about 55.0 arcmin or 0.92$\degr$. The data cube comprised spectra with 340 wavelength bins so that each sky pixel had a spectral bin size of 1.0 {\AA}, ranging from 1370 to 1710 {\AA}. We adopted the new effective areas of \citet{jo16}, derived after taking into account the degradation of detector sensitivity over time, to convert the count-rate unit (counts {\AA}$^{-1}$ s$^{-1}$) into the continuum unit (CU; photons cm$^{-2}$ sr$^{-1}$ {\AA}$^{-1}$ s$^{-1}$).

\subsection{Construction of the Diffuse Fluorescent H$_2$ Emission Map}

\begin{figure*}
	\begin{center}
		\includegraphics[height=7.5cm]{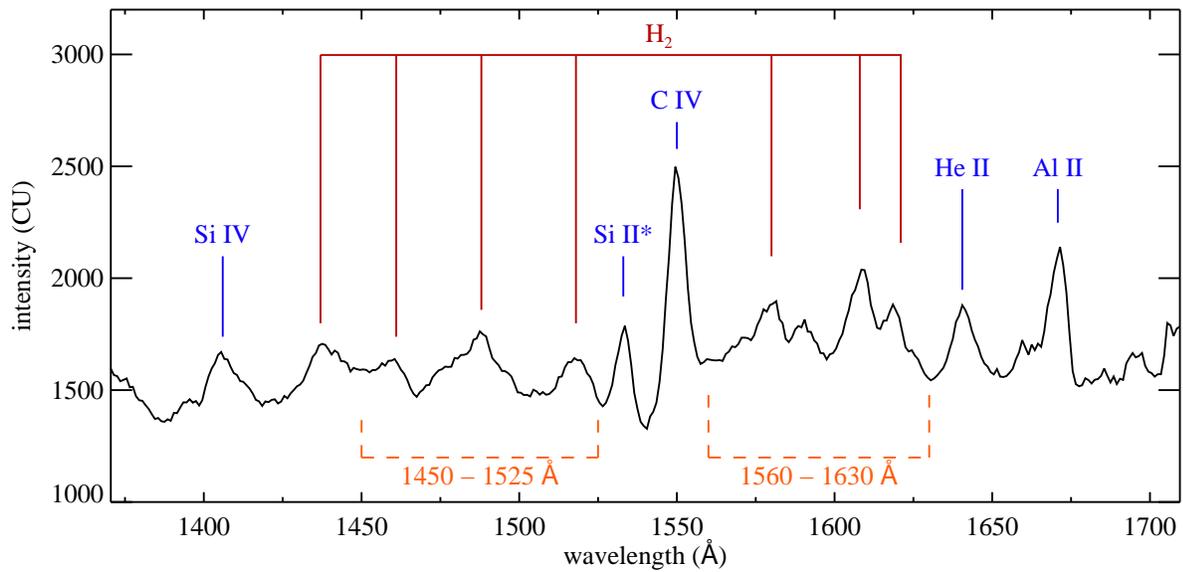}
	\end{center}
	\caption{
		Exposure time-weighted \textit{FIMS/SPEAR} \textit{L}-band spectrum. Several ion lines (\mbox{Si\,{\sc iv}}, \mbox{Si\,{\sc ii$^*$}}, \mbox{C\,{\sc iv}}, \mbox{He\,{\sc ii}}, and \mbox{Al\,{\sc ii}}) and the H$_2$ fluorescence emission features are identified. The orange dashed lines indicate the two bands in which the H$_2$ fluorescence emission features are dominant. \label{fig:allspec}}
\end{figure*}

Figure \ref{fig:allspec} presents the \textit{FIMS/SPEAR} all-sky exposure time-weighted diffuse background FUV spectrum with the direct stellar photons removed. This diffuse emission consists of various components: dust-scattered stellar continuum, hydrogen two-photon continuum, extragalactic background continuum, atomic emission lines, and H$_2$ fluorescence emission lines. Above the continuum background, atomic emission lines \mbox{Si\,{\sc iv}} $\lambda$1403, \mbox{Si\,{\sc ii$^*$}} $\lambda$1533, \mbox{C\,{\sc iv}} $\lambda\lambda$1548,1551, \mbox{He\,{\sc ii}} $\lambda$1640, and \mbox{Al\,{\sc ii}} $\lambda$1671 and several quasi-band-like features of the H$_2$ fluorescence emission lines are seen in the spectrum. The H$_2$ fluorescence features are dominant in two bands: 1450--1525 {\AA} and 1560--1630 {\AA}. The original data cube, with stellar photons removed, was rebinned to a larger wavelength bin size of 3 {\AA} to increase the SNR. Hence, the seven H$_2$ fluorescence emission lines with prominent peaks identified in Figure \ref{fig:allspec} are actually composed of many narrow lines but are observed as broad lines because of the coarse-grained spectrum.

It is essential to remove all the continuum background components as well as the atomic emission lines to extract only the H$_2$ fluorescence emission. The H$_2$ fluorescence emission map was constructed using a pixel size of $\sim$0.92$\degr$ ($N_{side}$ = 64). The spectrum of each pixel was obtained by smoothing the spectra of the pixels within a circle with weights proportional to the exposure time. The radius of the smoothing circle for each pixel was adaptively increased from 2$\degr$ up to 15$\degr$ in steps of 1$\degr$ until the SNR per spectral bin was $>$15. To extract the intensity of H$_2$ fluorescence emission, we first subtracted the continuum component from the spectrum smoothed over three adjacent spectral bins. The continuum spectrum was defined as line segments that connected local minima in the coarser spectral bins of 20 {\AA}, linearly interpolated at the original spectral bins, and smoothed by a boxcar ten spectral bins wide. While the present continuum subtraction algorithm is quiet primitive, we believe it works well as the smooth and monotonous continuum spectrum is expected without strong atomic lines at least in the wavelength regions where the H$_2$ emission features are dominant, as \citet{jo16} demonstrated for O-type and early B-type stars, which are the dominant sources of dust-scattered photons.

\begin{figure*}
	\begin{center}
		\includegraphics[height=7.5cm]{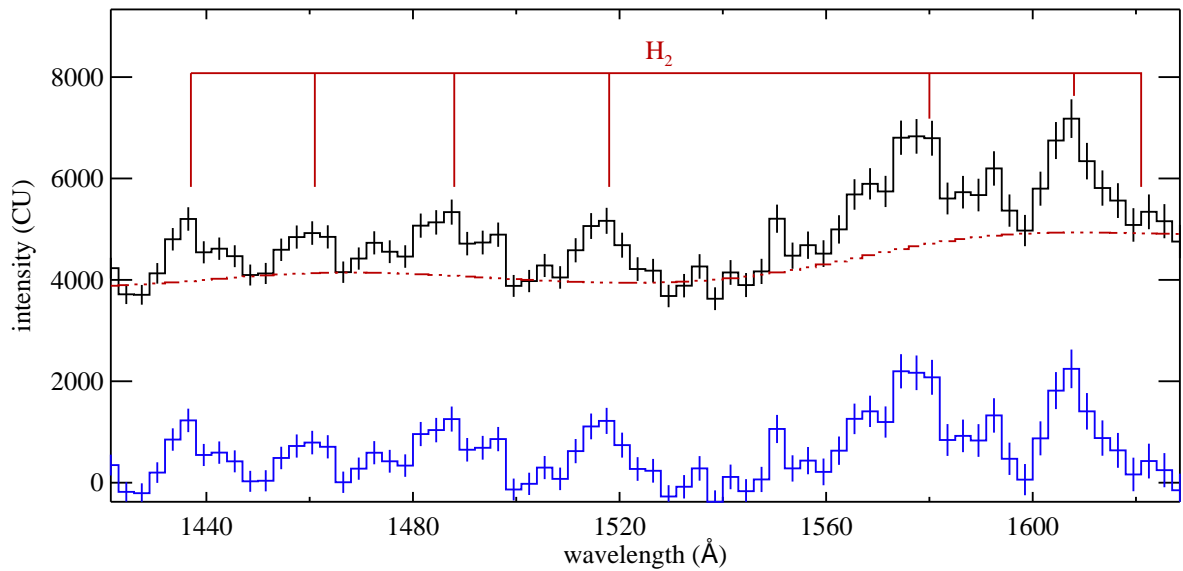}
	\end{center}
	\caption{
		Black solid line is an example of the exposure time-weighted FUV spectrum at the Galactic position of (\textit{l}, \textit{b}) $\sim$ (0$\degr$, 15$\degr$). The stellar continuum spectrum is indicated by a red dash-dotted line. The emission line spectrum (blue line), which includes the H$_2$ fluorescence emission lines, was obtained by subtracting the continuum spectrum from the FUV spectrum. Uncertainties are shown in both spectra by the vertical bars. \label{fig:samplespec}}
\end{figure*}

Figure \ref{fig:samplespec} shows an example of the measured spectrum and the constructed continuum spectrum for the Galactic coordinates (\textit{l}, \textit{b}) $\sim$ (0$\degr$, 15$\degr$), which was chosen arbitrarily among the regions where there are no nearby bright stars and whose H$_2$ fluorescence emission is not too bright. The smoothing radius used to obtain the spectrum was 4$\degr$. The spectral band shown in Figure \ref{fig:samplespec} includes not only the strong H$_2$ fluorescence emission lines but also the two atomic lines \mbox{Si\,{\sc ii$^*$}} $\lambda$1533 and \mbox{C\,{\sc iv}} $\lambda\lambda$1548,1551 which are rather weak in the present example. Hence, the total intensity of the H$_2$ fluorescence emission was obtained by integrating the emission line spectrum over the two bands (1450--1525 and 1560--1630 {\AA}), excluding the two atomic lines, and is presented in line units (LU: photons cm$^{-2}$ sr$^{-1}$ s$^{-1}$). We note that the wavelength band of 1450--1525 {\AA} includes the atomic line \mbox{N\,{\sc iv}} $\lambda$1486 which is significant only in hot gas regions such as supernova remnants. However, the contribution of this line to the total H$_2$ fluorescence emission intensity is generally less than about 13$\%$ even in hot gas regions, except in the Vela supernova remnant where it is as high as 30$\%$. The intensity of the FUV continuum (expressed in CU) was obtained by averaging the continuum spectrum over the spectral range of 1420--1630 {\AA}. We calculated the total intensity of the H$_2$ fluorescence emission and the FUV continuum intensity for each pixel to construct the all-sky map.

\begin{figure*}
	\begin{center}
		\includegraphics[trim=430 730 430 730,clip,height=15cm,angle=270]{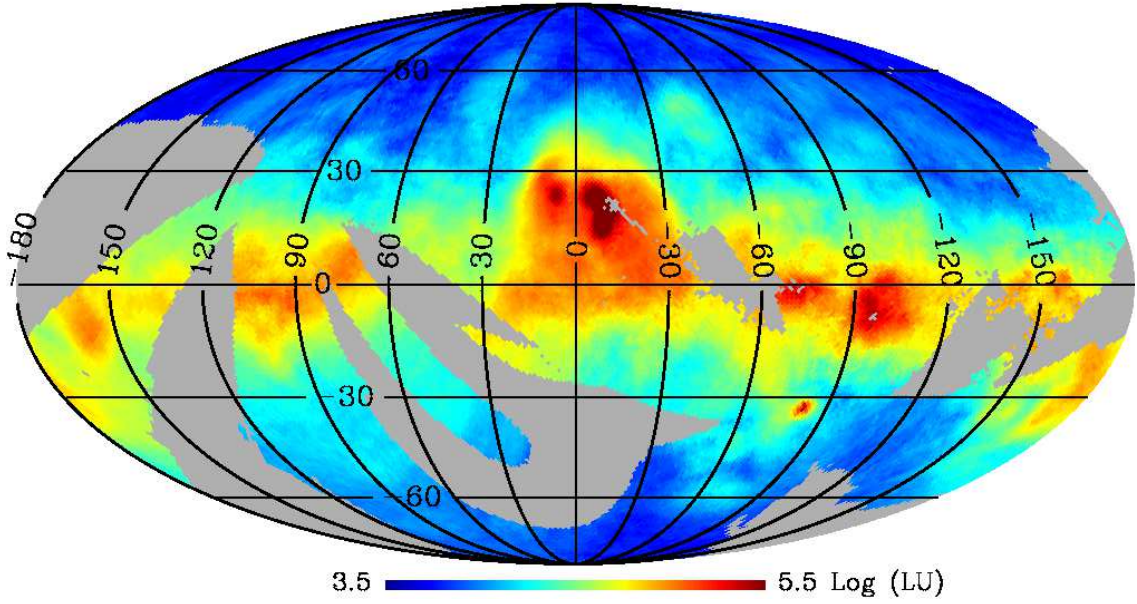}\\
		\vspace{10pt}
		\includegraphics[trim=430 730 430 730,clip,height=15cm,angle=270]{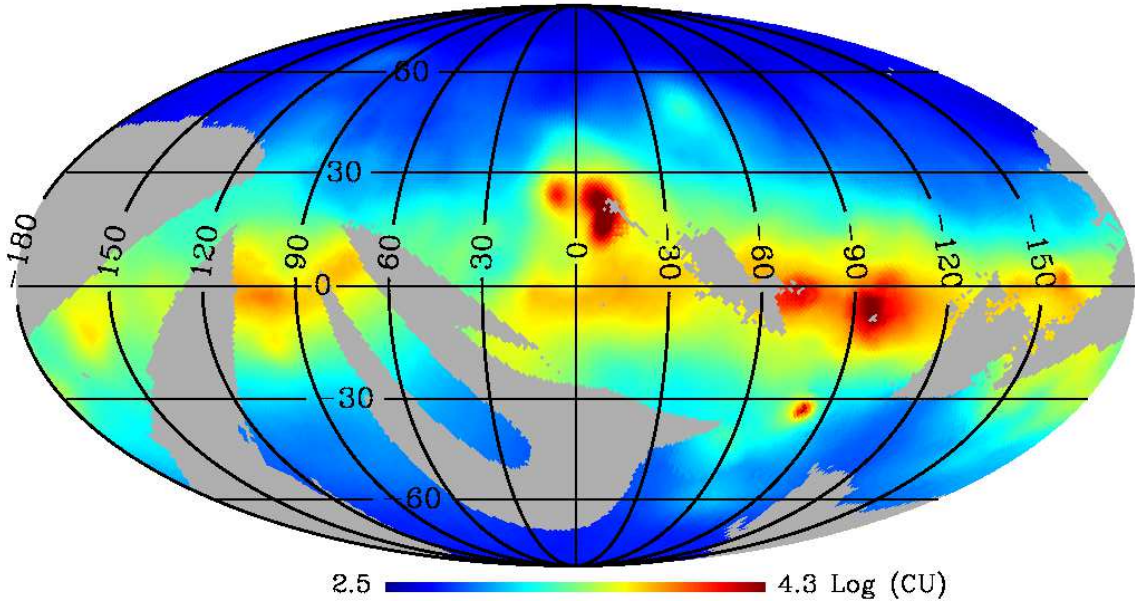}
	\end{center}
	\caption{
		(a) FUV H$_2$ fluorescence emission map, in line units (LU: photons cm$^{-2}$ sr$^{-1}$ s$^{-1}$). (b) FUV continuum map obtained after subtracting atomic lines and H$_2$ emission lines, in continuum units (CU: photons cm$^{-2}$ sr$^{-1}$ {\AA}$^{-1}$ s$^{1}$). (c) Map of the ratio of H$_2$ intensity to total FUV intensity, in percentage ($\%$). (d) Smoothing radius map, in units of degrees to achieve SNR $>$ 15 per spectral bin. Gray areas represent the pixels with zero exposure. \label{fig:h2map}}
\end{figure*}

\addtocounter{figure}{-1}
\begin{figure*}
	\begin{center}
		\includegraphics[trim=430 730 430 730,clip,height=15cm,angle=270]{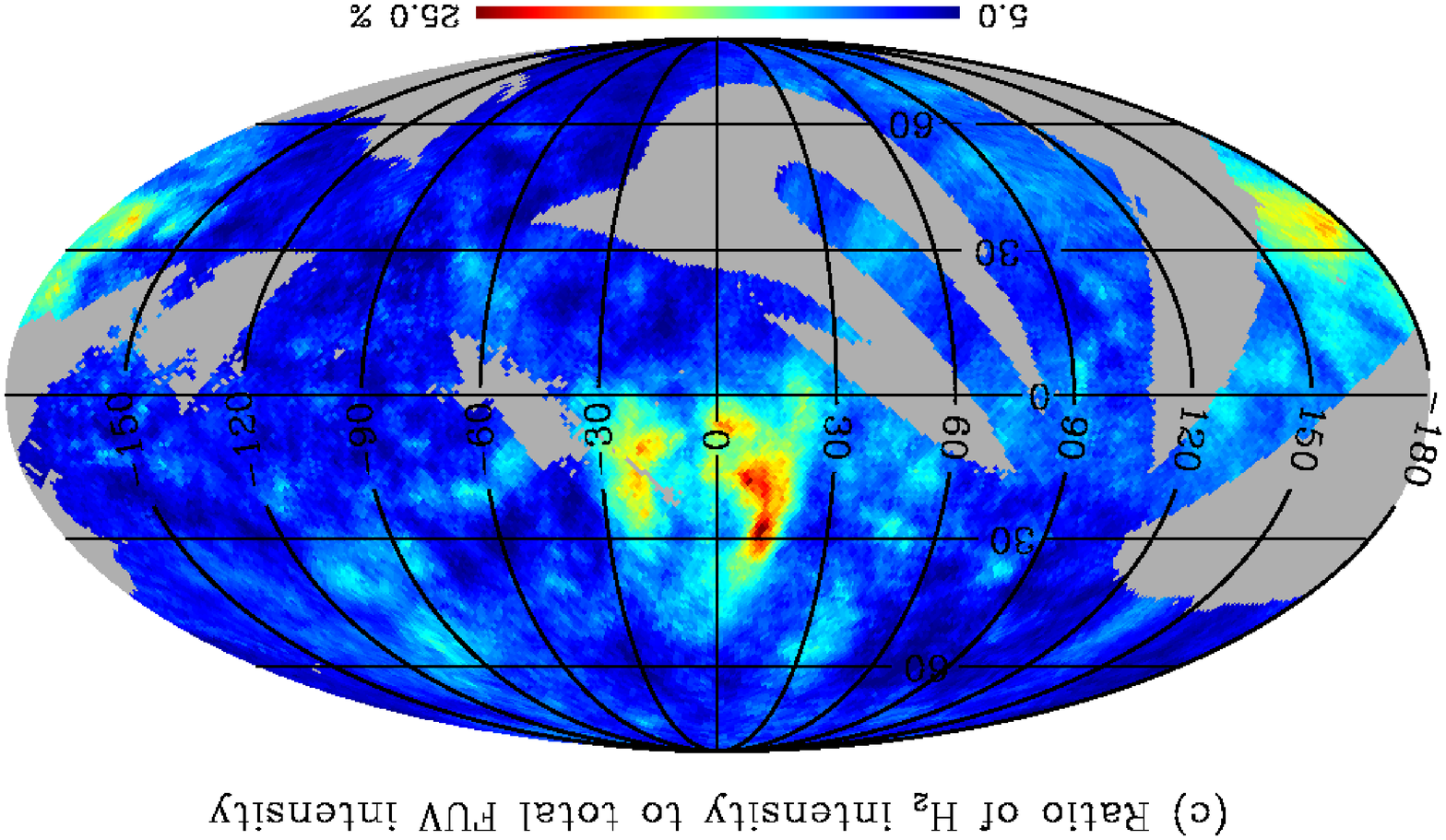}\\
		\vspace{10pt}
		\includegraphics[trim=430 730 430 730,clip,height=15cm,angle=270]{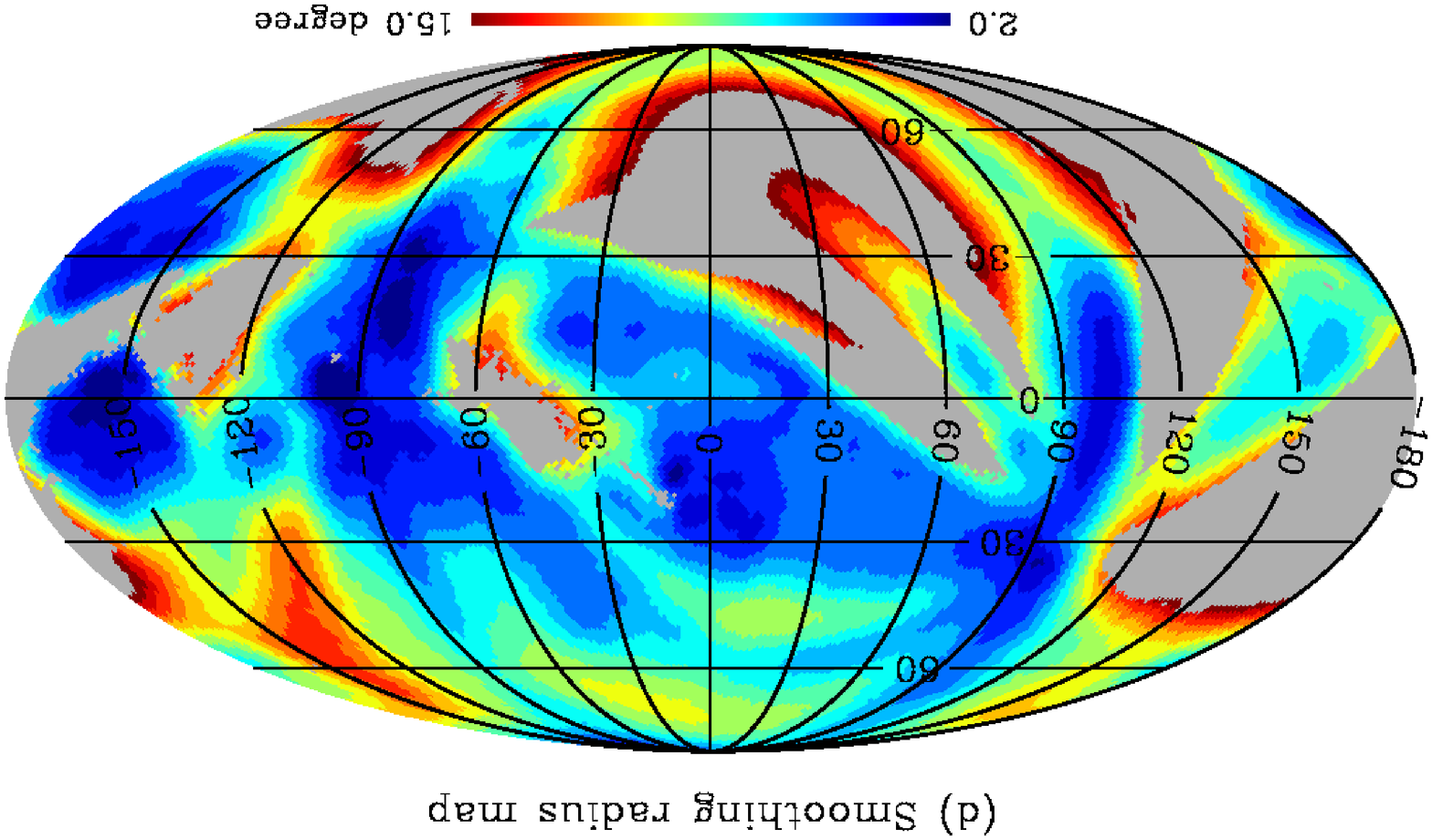}	
	\end{center}
	\caption{(continued)}
\end{figure*}

The resulting H$_2$ fluorescence emission map [Figure \ref{fig:h2map}(a)] is compared with the FUV continuum emission map [Figure \ref{fig:h2map}(b)] that originated mainly from the diffuse stellar light scattered by interstellar dust grains. Figures \ref{fig:h2map}(c) and (d) show the maps of the ratio of the H$_2$ intensity to the total FUV intensity and the ratio of the radii of the smoothing circles, respectively. The H$_2$ fluorescence emission map of Figure \ref{fig:h2map}(a) seems to display sharper structures than those in the FUV continuum emission map of Figure \ref{fig:h2map}(b) although the same smoothing procedure was applied to the two maps. The reason for this is that the H$_2$ fluorescence emission map was based on the line emissions that have much larger fluctuations across pixels than the FUV continuum. Nevertheless, the maps in Figures \ref{fig:h2map}(a) and (b) look very similar because both the intensity of the H$_2$ fluorescence emission and the intensity of the scattered emission of stellar light may be proportional to the strength of the incident UV radiation. The contribution of the H$_2$ fluorescence to the total FUV intensity, with a mean value of 8.7 $\pm$ 2.4$\%$ [depicted in Figure \ref{fig:h2map}(c)], does not vary significantly over the sky, except at the bright regions near the Ophiuchus cloud and the Eridanus region. The ratio of H$_2$ intensity to total FUV intensity increases up to about 26$\%$ near the Ophiuchus cloud at (\textit{l}, \textit{b}) $\sim$ (10$\degr$, 20$\degr$) and about 19$\%$ near the Eridanus region at (\textit{l}, \textit{b}) $\sim$ (190$\degr$, -40$\degr$). These two molecular clouds are known to have abundant H$_2$, i.e., \textit{N}(H$_2$) $\sim$ 10$^{22}$ cm$^{-2}$ for the Ophiuchus cloud \citep{cho15} and \textit{N}(H$_2$) $\sim$ 10$^{20}$ cm$^{-2}$ for the Eridanus region \citep{jo11}. It should be noted that these high-ratio regions of H$_2$ intensity to total FUV intensity are close to but displaced from the regions of high intensity in both H$_2$ fluorescence emission and continuum FUV. We believe the reason for this high H$_2$/continuum ratio comes from the geometry between the bright stars and the distributions of H$_2$ and dust in these regions because dust scattering strongly depends on the scattering angle with a peak in the forward direction, decreasing as the scattering angle increases, whereas H$_2$ fluorescence emission is isotropic. In fact, the high H$_2$/continuum FUV ratio regions have low continuum FUV intensities. The smoothing radius map [Figure \ref{fig:h2map}(d)], which was used to construct the H$_2$ fluorescence emission map, shows that the smoothing circles are large near the edges of the areas with zero exposure, an expected result because the number of pixels with nonzero exposure is small near these edges. The smoothing radius is small at locations where the exposure time and the photon count rate are high; the median value of the smoothing radius was $\sim$5$\degr$. In the following section, other all-sky maps are smoothed with the same smoothing radius map as that used for the H$_2$ fluorescence emission map.

\subsection{Comparison of the H$_2$ Fluorescent Emission Map with Other All-sky Maps}

\begin{figure*}
	\begin{center}
		\includegraphics[trim=430 730 430 730,clip,height=15cm,angle=270]{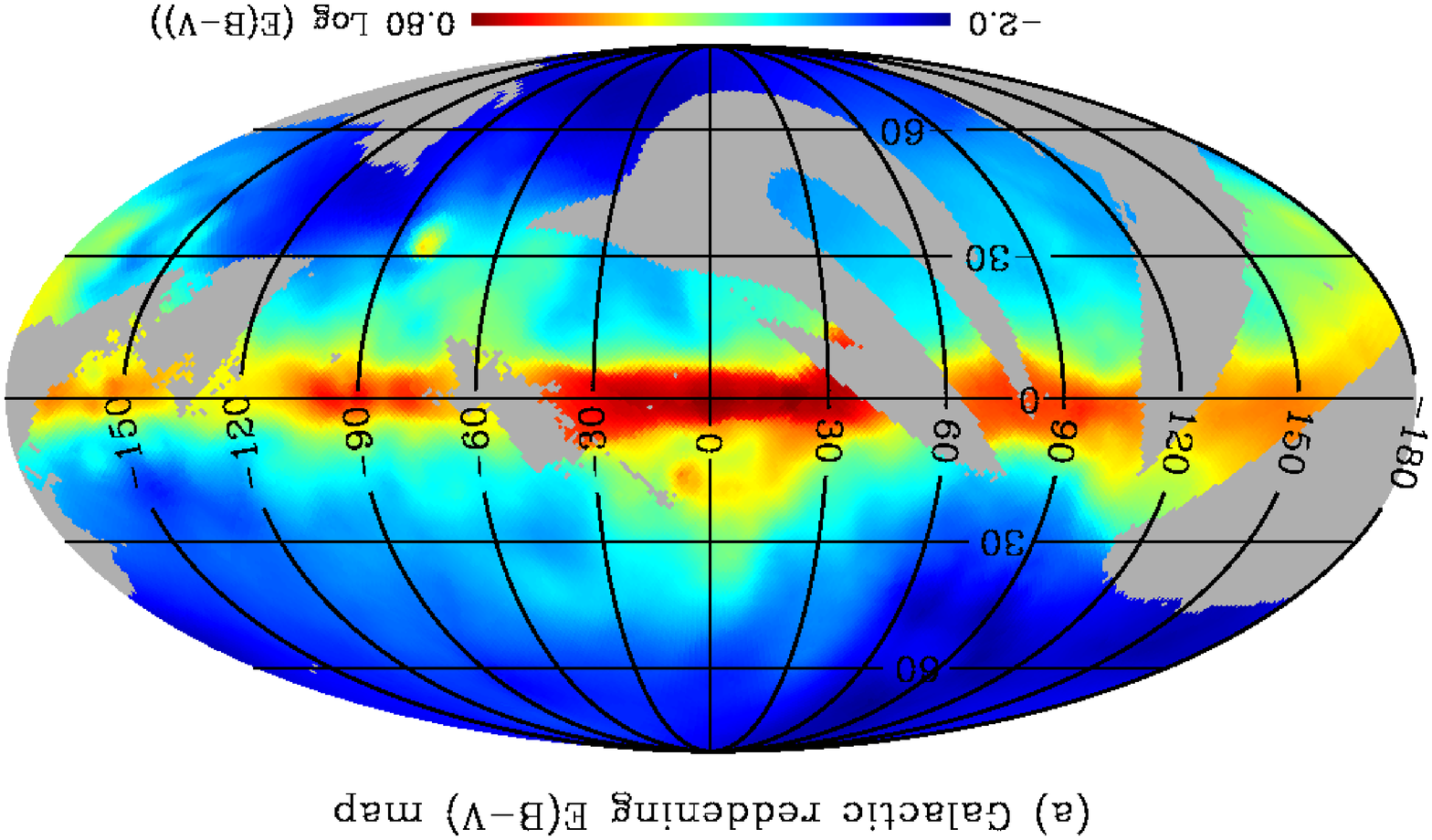}\\
		\vspace{10pt}
		\includegraphics[trim=430 730 430 730,clip,height=15cm,angle=270]{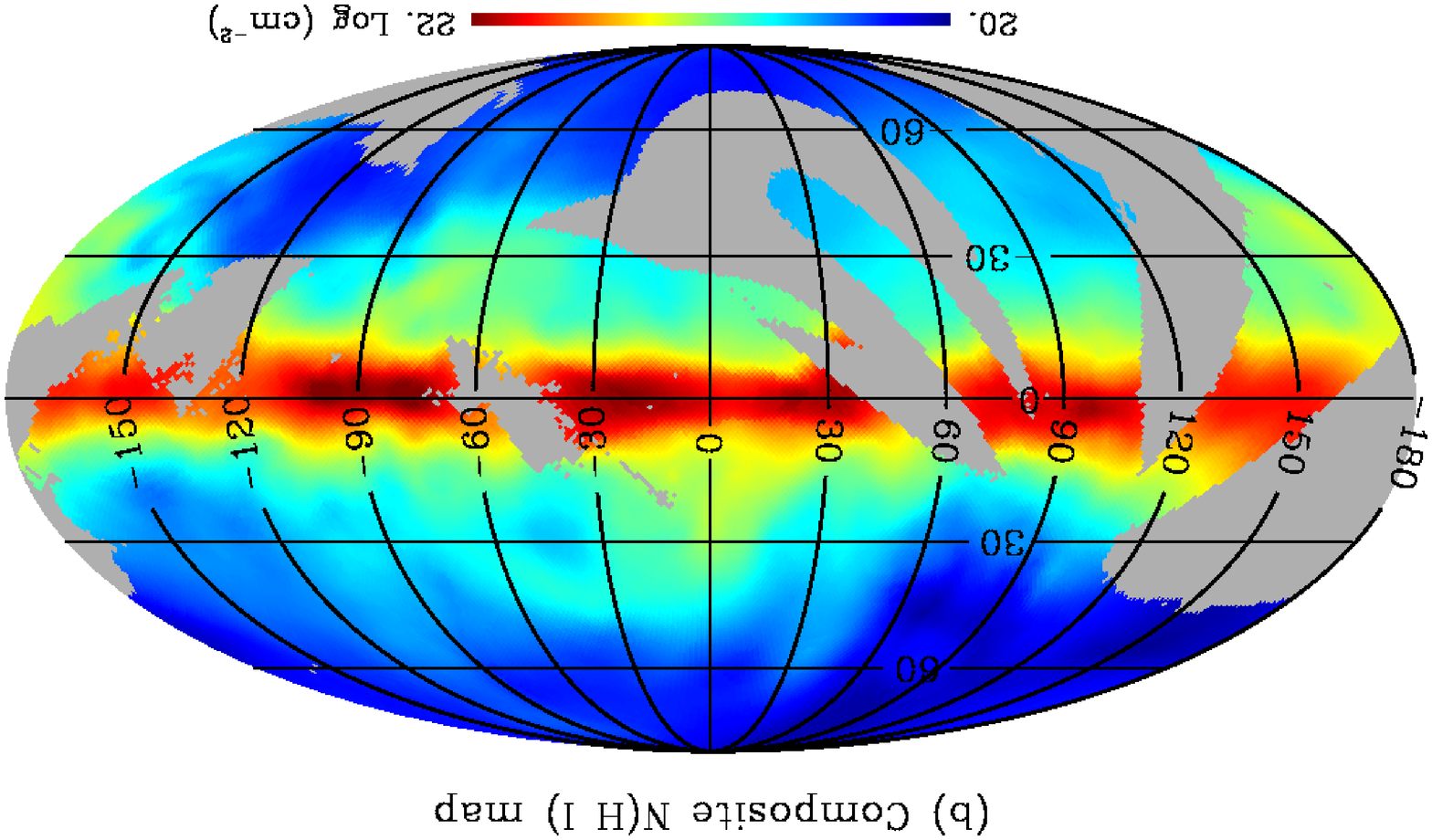}
	\end{center}
	\caption{
		(a) All-sky Galactic reddening \textit{E(B\textendash V)} map based on the reddening map of \citet{sch98}. (b) Composite all-sky map of \textit{N}(\mbox{H\,{\sc i}}) formed from the Leiden/Dwingeloo \mbox{H\,{\sc i}} survey data \citep{har97} and the composite \textit{N}(\mbox{H\,{\sc i}}) map of \citet{dic90}. (c) Composite all-sky H$\alpha$ intensity map generated by \citet{fin03}. (d) Planck all-sky CO \textit{J} = 1$\rightarrow$0 line map \citep{ade14}. \label{fig:othermap}}
\end{figure*}

\addtocounter{figure}{-1}
\begin{figure*}
	\begin{center}
		\includegraphics[trim=430 730 430 730,clip,height=15cm,angle=270]{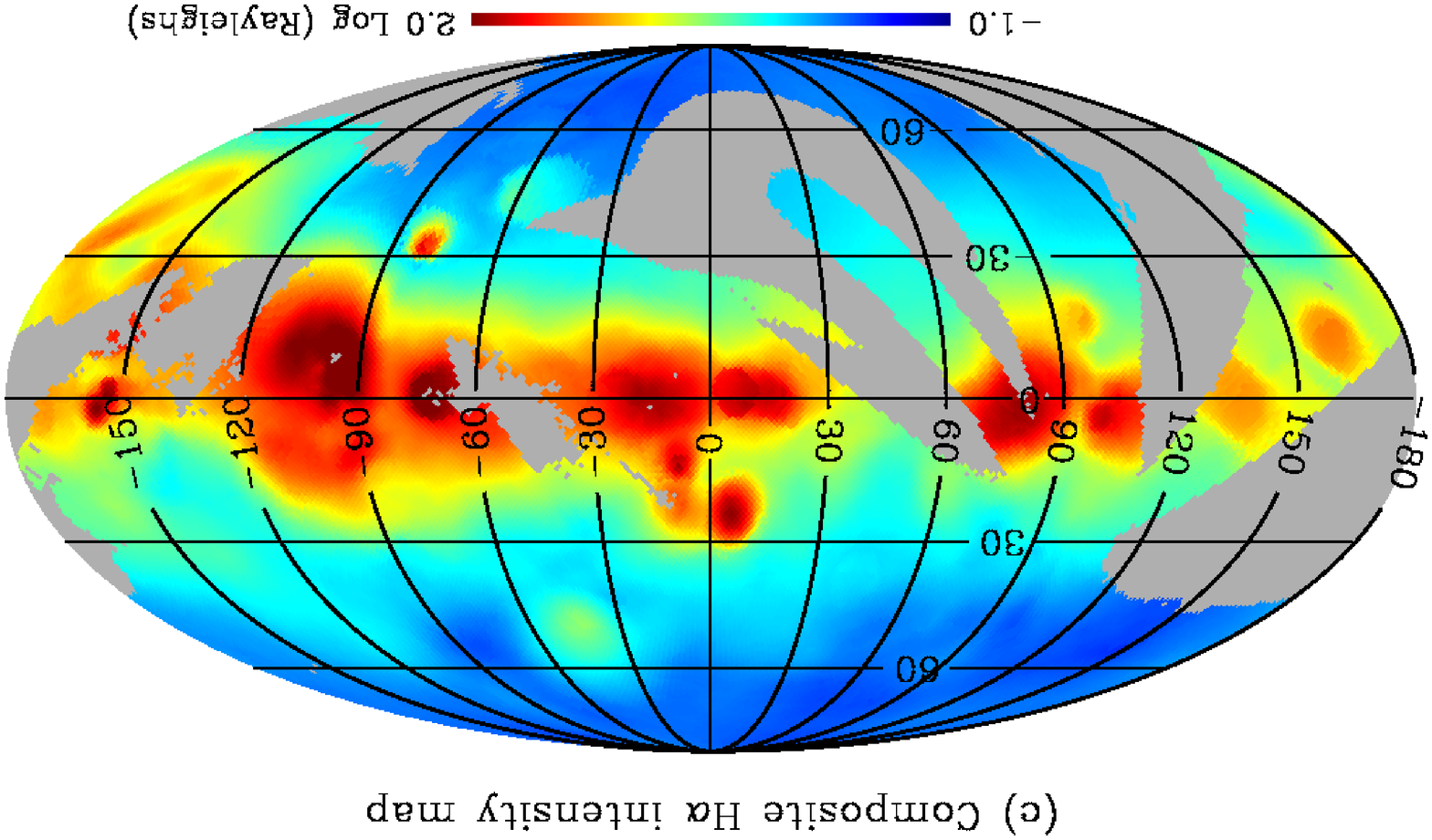}\\
		\vspace{10pt}
		\includegraphics[trim=430 730 430 730,clip,height=15cm,angle=270]{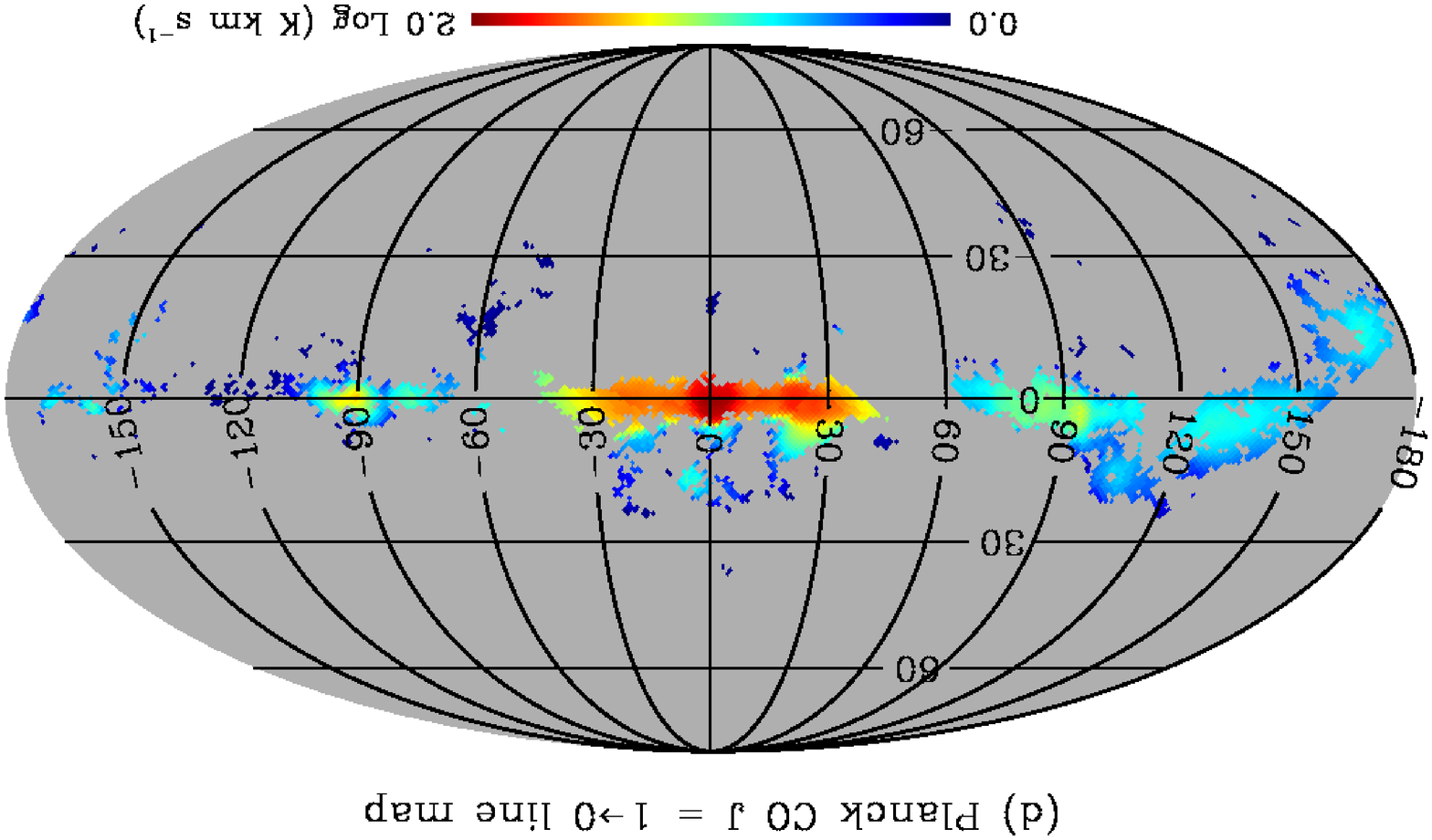}
	\end{center}
	\caption{(continued)}
\end{figure*}

\begin{figure*}
	\begin{center}
		\includegraphics[height=7.5cm]{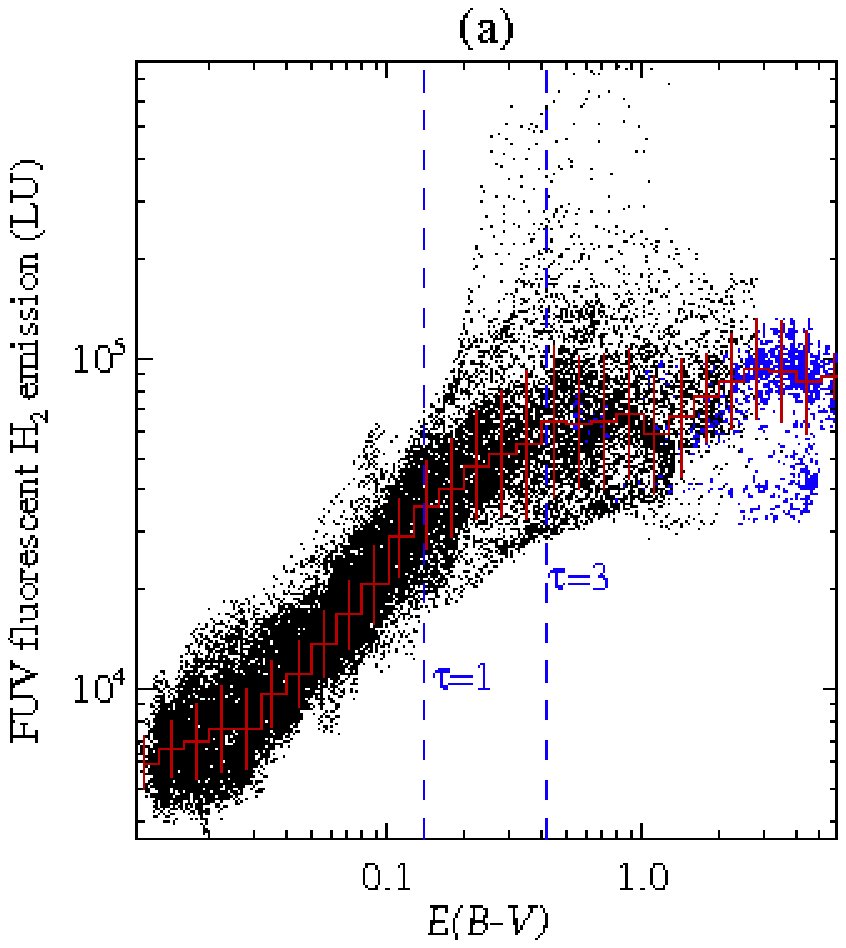}\hspace{10pt}
		\includegraphics[height=7.5cm]{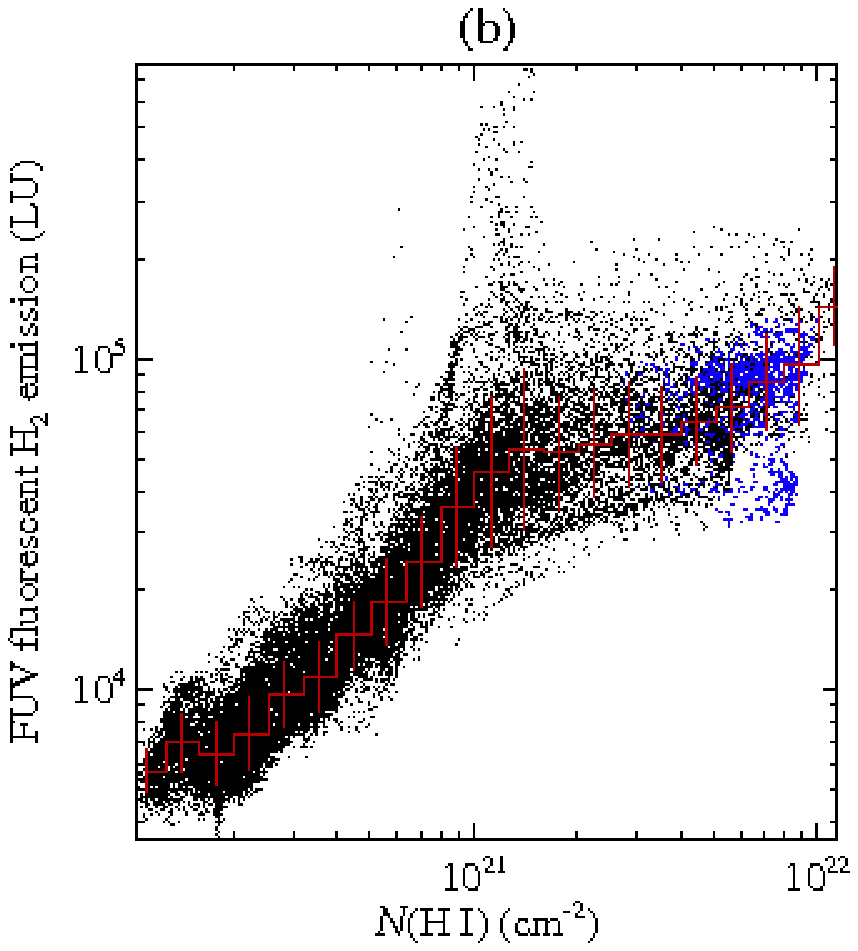}\\
		\vspace{10pt}
		\includegraphics[height=7.5cm]{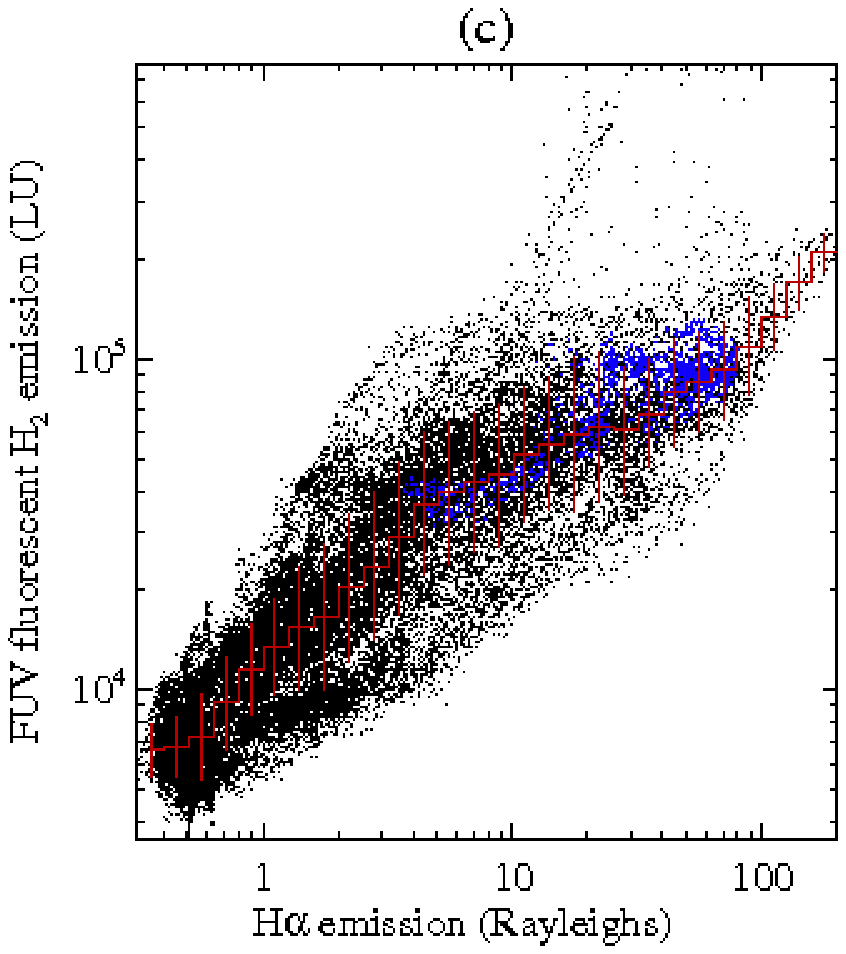}\hspace{10pt}
		\includegraphics[height=7.5cm]{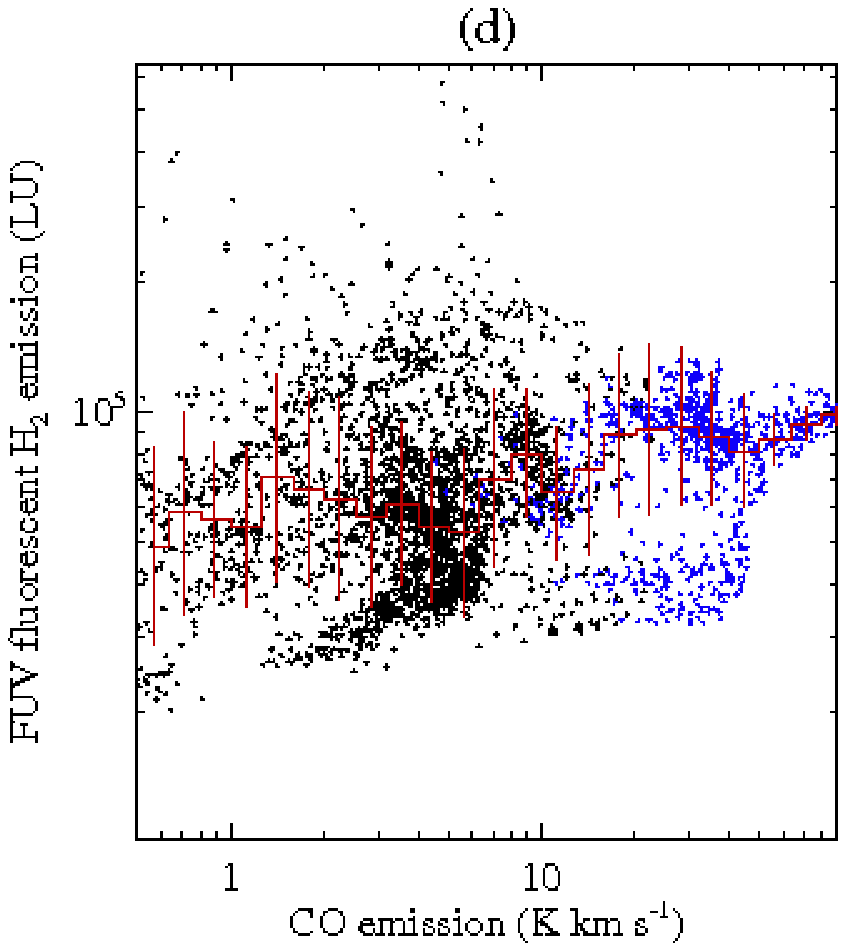}
	\end{center}
	\caption{
		Correlation of the FUV fluorescent H$_2$ emission map with all-sky maps of (a) \textit{E(B\textendash V)}, (b) \textit{N}(\mbox{H\,{\sc i}}), (c) H$\alpha$, and (d) CO. The blue data points were obtained from the region around the Galactic center with $\left| b\right|  <$ 5$\degr$ and $\left| l\right|  <$ 60$\degr$, and the black data points were obtained from the rest of the sky. \label{fig:pixel1}}
\end{figure*}

We compared the FUV fluorescent H$_2$ emission map with other all-sky maps, especially maps of interstellar dust extinction, \mbox{H\,{\sc i}}, and CO, which may be closely related to H$_2$, to examine any correlations between these maps. For example, the atomic hydrogen density is expected to be proportional to the square root of the molecular hydrogen density in a steady state with constant photodissociation and recombination rates \citep{dra11}, as is the case for the FUV intensity of fluorescent H$_2$ emission in the optically thin limit \citep{ste89}. This implies that the FUV fluorescent H$_2$ emission should be proportional to the column density of \mbox{H\,{\sc i}} as well as \textit{E(B\textendash V)}, which is generally proportional to the column density of \mbox{H\,{\sc i}}. We also compared the FUV fluorescent H$_2$ emission map with the H$\alpha$ map because the UV radiation field that excites the H$_2$ fluorescence emission also ionizes H to produce H$\alpha$ emission. We downloaded the \textit{E(B\textendash V)}, \mbox{H\,{\sc i}} column density [\textit{N}(\mbox{H\,{\sc i}})], and H$\alpha$ intensity all-sky maps from the Legacy Archive for Microwave Background Data Analysis.$\footnote[1]{http://lambda.gsfc.nasa.gov/product/foreground/f\_products.cfm}$ The Planck all-sky CO \textit{J} = 1$\rightarrow$0 line map was downloaded from the NASA/IPAC Infrared Science Archive \citep{pla16}.$\footnote[2]{http://irsa.ipac.caltech.edu/data/Planck/release\_2/all-sky-maps/maps/component-maps/foregrounds/HFI\_CompMap\_CO-Type2\_2048\_R2.00.fits}$ All four maps were created with the same resolution parameter ($N_{side}$ = 64) used for the H$_2$ fluorescence map and were smoothed using the smoothing radius map in Figure \ref{fig:h2map}(d) after masking out the areas with zero exposure, as was done for the H$_2$ fluorescence map. The resulting four maps are shown in Figure \ref{fig:othermap}. Regions where the SNR $<$ 5.0 were excluded from the Planck CO map. Only $\sim$7$\%$ of the CO map near the Galactic plane remained after the masking and smoothing procedures were performed.

Figure \ref{fig:pixel1} compares the H$_2$ fluorescence emission map of Figure \ref{fig:h2map}(a) with the four all-sky maps of Figure \ref{fig:othermap} on a logarithmic scale. The blue data points were obtained from the region around the Galactic center with $\left| b\right|  <$ 5$\degr$ and $\left| l\right|  <$ 60$\degr$, and the black points were obtained from the rest of the sky. In general, the H$_2$ fluorescence intensity correlates well with \textit{E(B\textendash V)}, \textit{N}(\mbox{H\,{\sc i}}), and H$\alpha$ for the black data points, especially in the low-intensity regions (i.e., optically thin, high latitudes), while it becomes saturated and shows a large scatter in the optically thick regions including the region around the Galactic center. The blue dashed lines in Figure \ref{fig:pixel1}(a) indicate the reddening values that correspond to optical depths at 1550 {\AA} ($\tau_{1550}$) of 1 and 3. The linear proportionality of the H$_2$ fluorescence intensity disintegrates at optically thick regions where $\tau_{1550} >$ 1. Similarly, the intensity of the FUV fluorescent H$_2$ emission is no longer proportional to \textit{N}(\mbox{H\,{\sc i}}) above \textit{N}(\mbox{H\,{\sc i}}) $\sim$ 10$^{21}$ cm$^{-2}$, which corresponds to $\tau_{1550} \sim$ 1, as seen in Figure \ref{fig:pixel1}(b). Figure \ref{fig:pixel1}(c) shows that the H$_2$ fluorescence emission is also correlated with the H$\alpha$ emission, which is expected because they have common radiation sources of O- and B-type stars. H$\alpha$ photons are less attenuated by interstellar dust than FUV photons: the optical depth of H$\alpha$ photons is about three times less than that of FUV photons. Hence, the range of H$\alpha$ intensity is wider than that of FUV intensity. Finally, the CO intensity shows a fairly poor correlation with FUV fluorescent H$_2$ intensity, especially for the blue data points obtained in the region around the Galactic center. Nevertheless, the black data points seem to show a correlation of CO intensity with the FUV fluorescent H$_2$ emission, though they are still scattered significantly.

\begin{figure*}
	\begin{center}
		\includegraphics[trim=430 730 430 730,clip,height=15cm,angle=270]{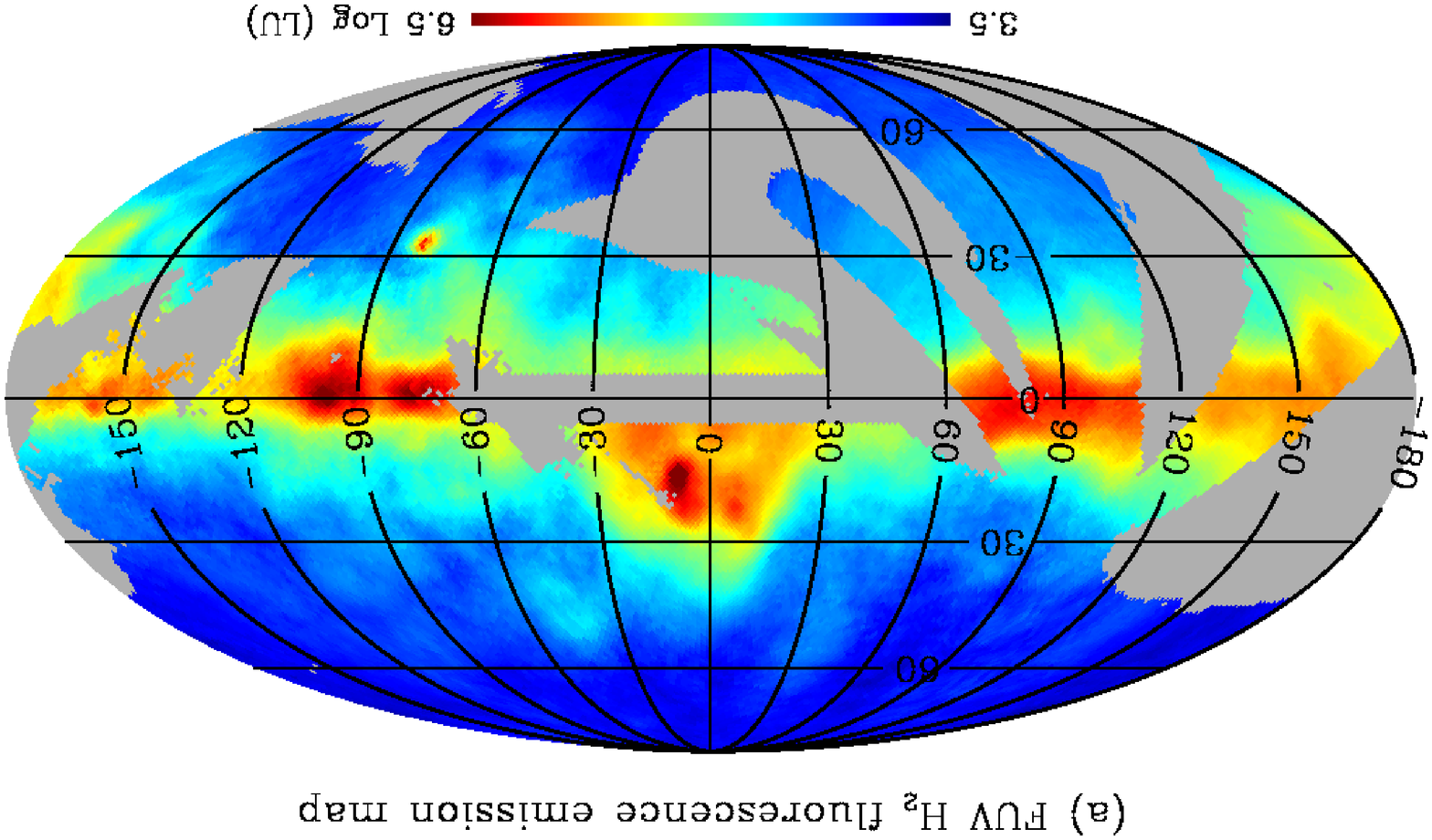}\\
		\vspace{10pt}
		\includegraphics[trim=430 730 430 730,clip,height=15cm,angle=270]{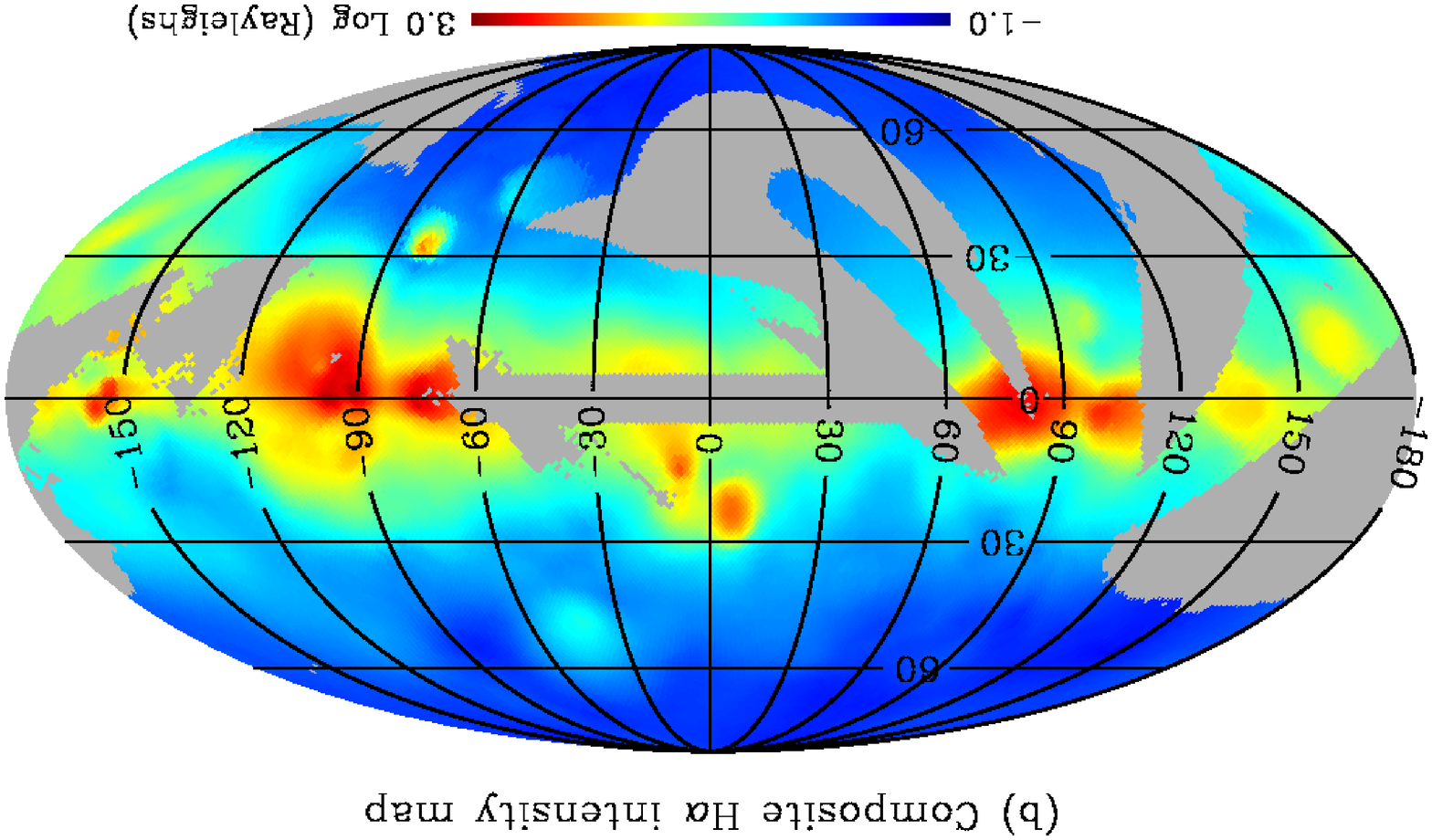}
	\end{center}
	\caption{
		(a) FUV fluorescent H$_2$ emission map and (b) H$\alpha$ map after correcting for dust extinction. The strongly extinct region of $\left| l\right|  <$ 60$\degr$ and $\left| b\right|  <$ 5$\degr$ around the Galactic center is excluded in the figures. \label{fig:extcorr}}
\end{figure*}

\begin{figure*}
	\begin{center}
		\includegraphics[height=7.5cm]{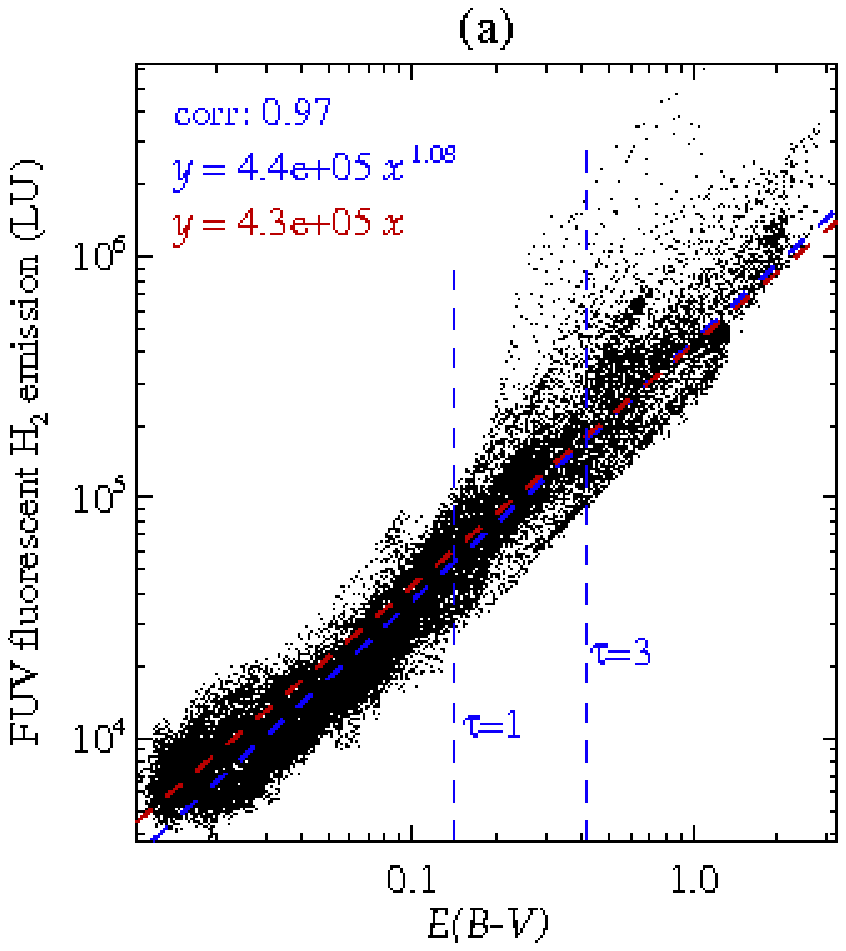}\hspace{10pt}
		\includegraphics[height=7.5cm]{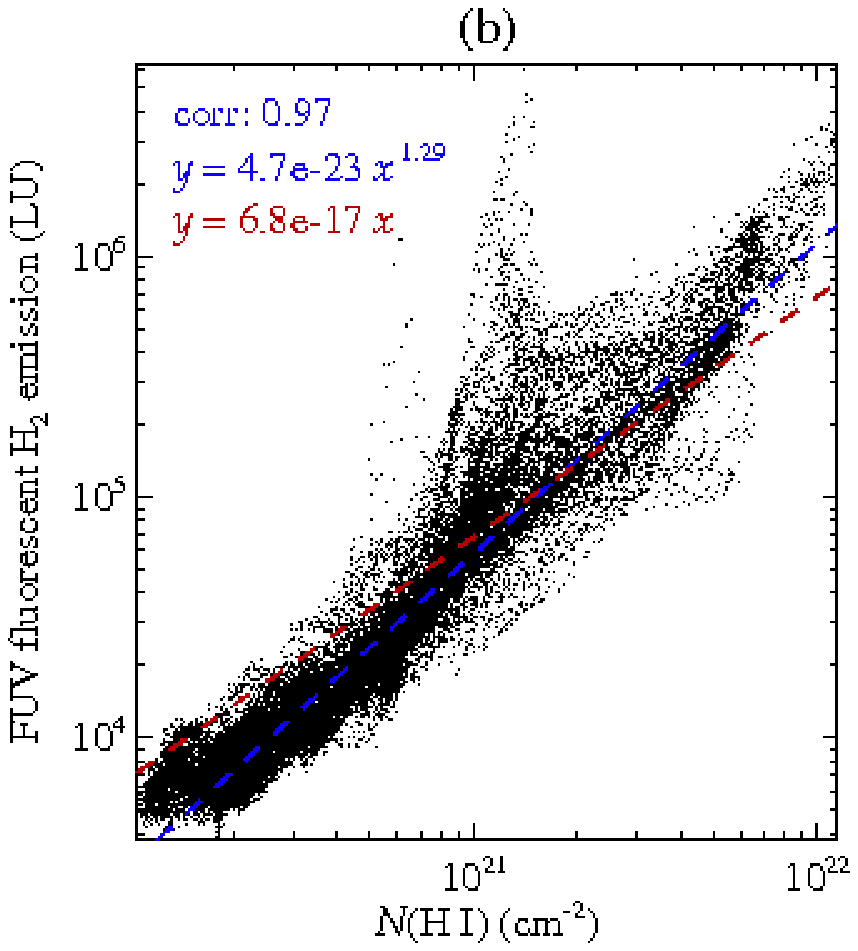}\\
		\vspace{10pt}
		\includegraphics[height=7.5cm]{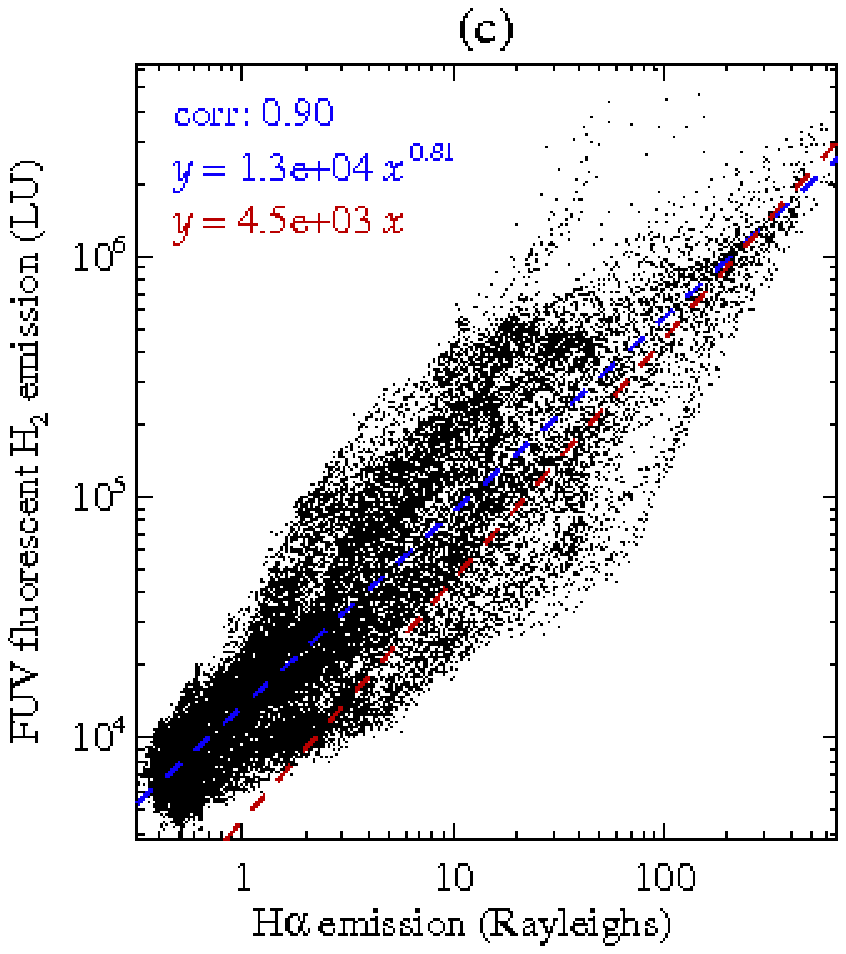}\hspace{10pt}
		\includegraphics[height=7.5cm]{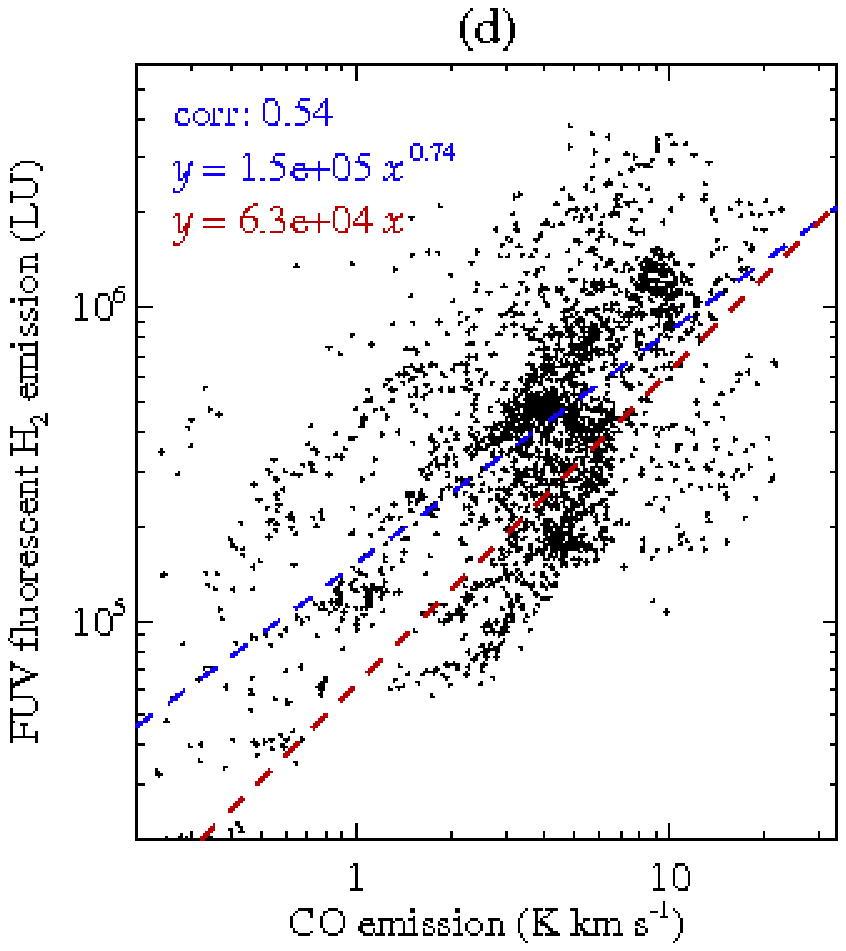}
	\end{center}
	\caption{
		Correlation of the FUV fluorescent H$_2$ emission with (a) \textit{E(B\textendash V)}, (b) \textit{N}(\mbox{H\,{\sc i}}), (c) H$\alpha$, and (d) CO after correcting H$_2$ fluorescence emission and H$\alpha$ emission for dust extinction. Best-fit equations and Spearman's rank correlation coefficient are shown in the upper left of each panel. \label{fig:pixel2}}
\end{figure*}

We corrected for the effects of dust extinction on the FUV and H$\alpha$ intensities to find the true relationship between the H$_2$ fluorescence emission map and the other maps, while the effects of dust extinction on the CO 2.6-mm and the \mbox{H\,{\sc i}} 21-cm lines were negligible. We assumed that the emission sources and the interstellar dust are uniformly mixed, so the extinction-corrected intensity is given by $I\times\tau/(1-e^{-\tau})$, where \textit{I} is the observed intensity. If we use the Milky Way extinction curve of \citet{wei01}, where \textit{R}$_V$ = 3.1, then the optical depths at FUV ($\sim$1550 {\AA}) and H$\alpha$ ($\sim$6562 {\AA}) emission in terms of \textit{E(B\textendash V)} are $\tau_{1550} = 7.30 \times \textit{E(B\textendash V)}$ and $\tau_{6562} = 2.22\times\textit{E(B\textendash V)}$, respectively. The original \textit{E(B\textendash V)} map, whose resolution is approximately 9.5 arcmin, and the H$\alpha$ intensity map, whose resolution is approximately 6 arcmin, were re-binned to have the same pixel size of about 55.0 arcmin as that of the H$_2$ fluorescence map. Before applying extinction correction, we masked the region around the Galactic center of $\left| l\right|  <$ 60$\degr$ and $\left| b\right|  <$ 5$\degr$ because the above extinction formula that assumes a uniform mixing of emission sources and interstellar dust may especially be invalid in the Galactic center region. Smoothing was performed excluding this region around the Galactic center using the smoothing radius map of Figure \ref{fig:h2map}(d). The results are displayed in Figure \ref{fig:extcorr}. The intensity of FUV emission is observed to increase by a factor of about two, on average, after correcting for dust extinction, while the H$\alpha$ intensity increases by about 30$\%$. The FUV intensity at (\textit{l}, \textit{b}) $\sim$ (0$\degr$, 15$\degr$), shown as an example in Figure \ref{fig:samplespec}, increases by a factor of about four.

Figure \ref{fig:pixel2} presents the four correlation plots shown in Figure \ref{fig:pixel1} obtained using the extinction-corrected maps. The red and oblique blue dashed lines indicate the best-fit power law and linear function, respectively, passing through the origin, to the correlation plots. All four parameters show strong correlations with the H$_2$ fluorescence emission, with the correlations of dust extinction and \textit{N}(\mbox{H\,{\sc i}}) being especially strong. The large scatter in Figure \ref{fig:pixel2}(c) may have been caused by the different mechanisms that produce the H$_2$ fluorescence emission and the H$\alpha$ emission; the latter is produced from ionized hydrogen and the former is produced from cold and dense molecular clouds. Although there is still a large amount of scatter, an improved correlation is observed between the extinction-corrected H$_2$ fluorescence emission and CO emission. We note that the de-reddening procedure performed for the FUV fluorescent H$_2$ intensity itself enhances the correlation with CO as well as \textit{E(B\textendash V)} and \mbox{H\,{\sc i}}. Hence, the widely scattered CO data will not be used in the following discussions.


\section{Discussion}

\subsection{Photodissociation Region Modeling}

Because fluorescent H$_2$ emission should be associated with H$_2$ column density, we modeled the H$_2$ fluorescence-emitting regions as a PDR using the plane-parallel PDR code CLOUD \citep{bla87,van86} to investigate the spatial distribution of H$_2$ in the Milky Way. CLOUD calculated all the narrow H$_2$ fluorescence emission lines, which we re-binned with a resolution of 3 {\AA} for comparison with the observed data. The main input parameters of the CLOUD code were (1) the sticking probability and formation efficiency of H$_2$ on dust grains ($y_F$) \citep{bla73a,bla73b,bla76}, (2) hydrogen density ($n_H$), (3) cloud temperature (\textit{T}), (4) strength of incident UV radiation ($I_{UV}$) in units of the average UV radiation field of \citet{dra78}, and (5) \textit{N}(H$_2$). The H$_2$ fluorescence emission spectrum for a line of sight is the integration of fluorescence spectra from clumpy gas clouds located along the line of sight in diverse environments. However, because there are limitations in determining all the physical parameters of the individual gas clouds for each line of sight simultaneously, we assumed a single-cloud model along each line of sight. In the optically thin limit with the extinction effect ignored, this single cloud assumption overestimates the molecular hydrogen column density by a factor of the number of clouds that actually lie along the line of sight. The difference should be much smaller in reality because of the extinction effect. In addition, some parameters that are less important in the PDR models were fixed to their typical values.

In the PDR calculation for a gas cloud, the total flux of the FUV fluorescent H$_2$ emission is most sensitive to variations in $I_{UV}$ and \textit{N}(H$_2$). In contrast, the total flux of the FUV fluorescent H$_2$ emission does not change significantly when other parameters vary, as long as the pressure in the gas cloud is kept at a typical gas pressure value for the ISM (\textit{p}/\textit{k} $\sim$ 1000 cm$^{-3}$ K). Therefore, the three least-important parameters were fixed at typical values of $y_F$ = 1, $n_H$ = 10 cm$^{-3}$, and \textit{T} = 100 K, which correspond to \textit{p}/\textit{k} = 1000 cm$^{-3}$ K. Thus, the H$_2$ formation rate coefficient was 2.4 $\times$ 10$^{-17}$ cm$^{3}$ s$^{-1}$ K. We adopted dust grain model 2 of \citet{bla87}, in which the scattering albedo and the asymmetry phase factor at 1000 {\AA} are 0.6 and 0.5, respectively [see \citet{bla87} for more details about the dust grain model]. The assumption about the three least-important parameters may slightly bias the results, but our results did not change significantly when $y_F$, $n_H$, and \textit{T} were varied within reasonable ranges.

\begin{figure*}
	\begin{center}
		\includegraphics[height=7.5cm]{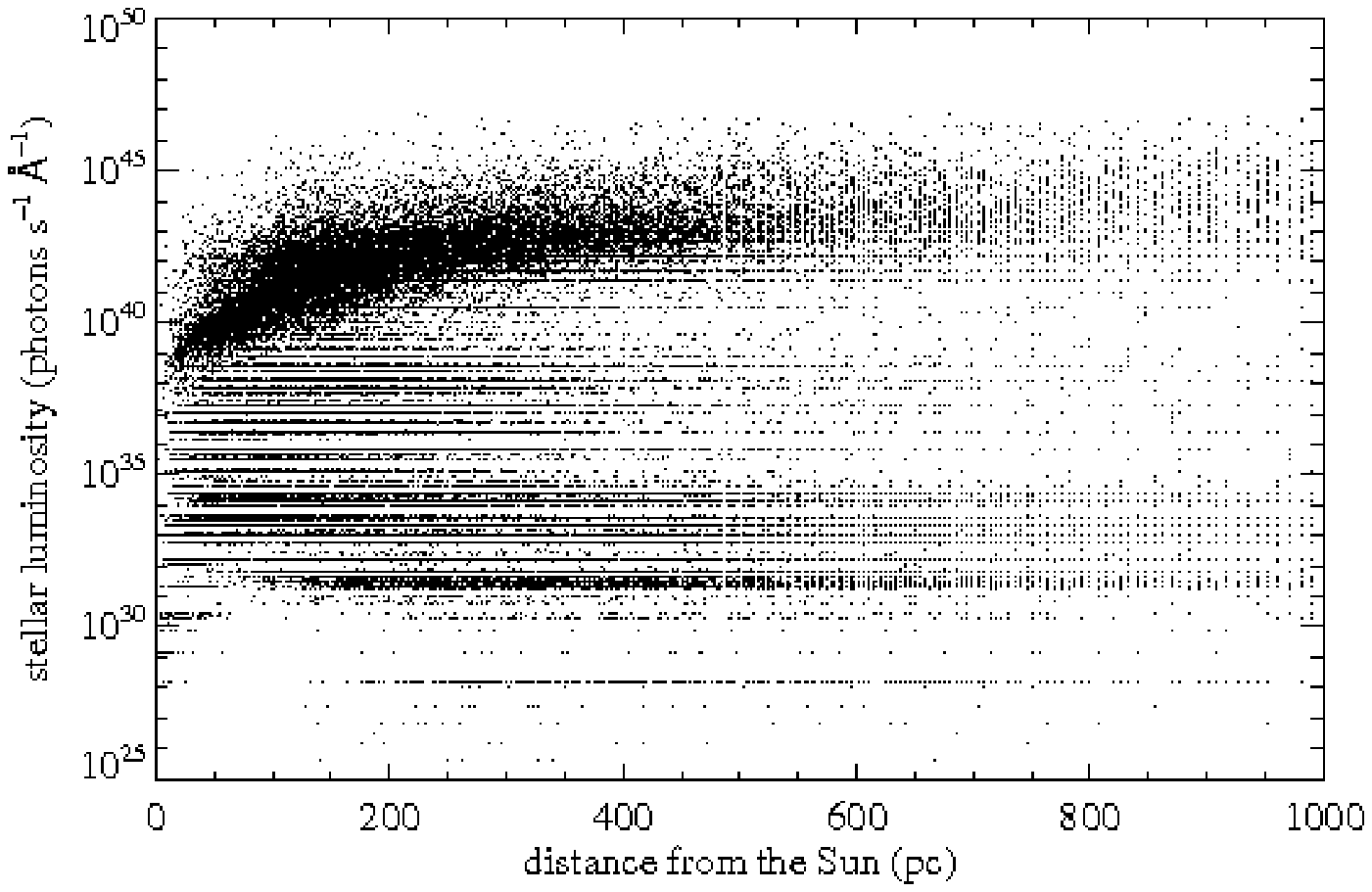}
	\end{center}
	\caption{
		Scatterplot of stellar luminosity as a function of distance derived from the TD1 and Hipparcos star catalogs for 87,863 stars used in this study. \label{fig:star}}
\end{figure*}

With our simplified model, the intensity of the FUV fluorescent H$_2$ emission depends on only $I_{UV}$ and \textit{N}(H$_2$), and \textit{N}(H$_2$) can be obtained if we determine $I_{UV}$, which is relatively well constrained. We used the TD1 star catalog \citep{tho78}, which provides the observed FUV fluxes of stars, and the Hipparcos star catalog \citep{per97}, which provides parallaxes and spectral types of stars, to calculate stellar luminosities in 3D space. The observed FUV fluxes of the TD1 stars were converted into luminosities by accounting for the effect of distance and dust extinction. The luminosities of the Hipparcos stars without TD1 flux information were calculated using the Kurucz stellar models \citep{cas04} and the stellar spectral types. Figure \ref{fig:star} shows a scatterplot of the FUV luminosity at 1565 {\AA} as a function of distance derived from the catalogs for the 87,863 stars used in this study. The dense population above approximately 10$^{38}$ photons s$^{-1}$ {\AA}$^{-1}$ corresponds to the TD1 stars (21,181 stars in total) and the approximately uniform background corresponds to the stars of the Hipparcos catalog (66,682 stars in total). The bimodality observed in the scatterplot is due to the difference in wavelength range that the two catalogs are based on. The Hipparcos catalog was produced with optical observations and includes late-type stars that were not observed by TD1 because of their faint luminosity in the FUV. The appearance of horizontal and vertical ``stripes'' in the figure is due to the discreteness of the stellar spectral type and parallax, respectively, of the Hipparcos catalog. The dense population of the TD1 stars indicates that fewer stars were observed as the distance from the Sun increased because of the effects of dust extinction and the inverse square law. However, the median luminosity slightly increases with distance from the Sun because brighter stars can be observed more easily than faint stars when they are at large distances from the Sun. We calculated the 3D mean intensity $I_{UV}$, averaged over a solid angle, at arbitrary locations around the Sun using the stellar luminosities and stellar locations. As the distance from the Sun increases, the 3D mean intensity $I_{UV}$ may become inaccurate.

\begin{figure*}
	\begin{center}
		\includegraphics[height=7.5cm]{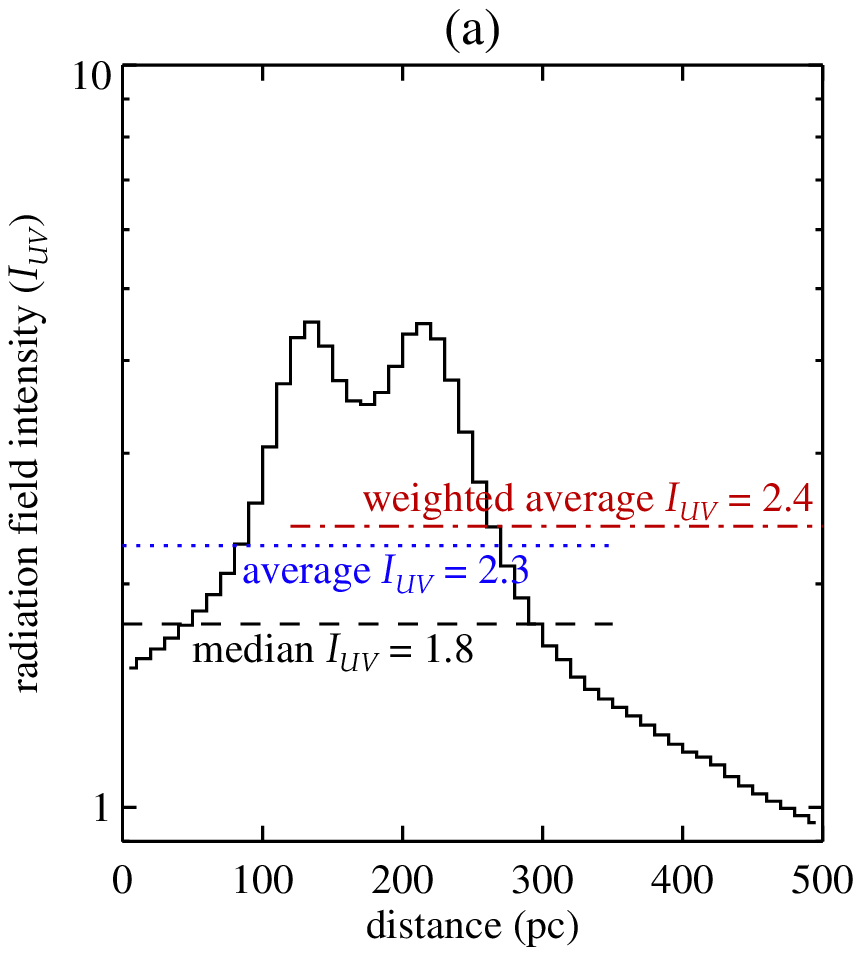}\hspace{10pt}
		\includegraphics[height=7.5cm]{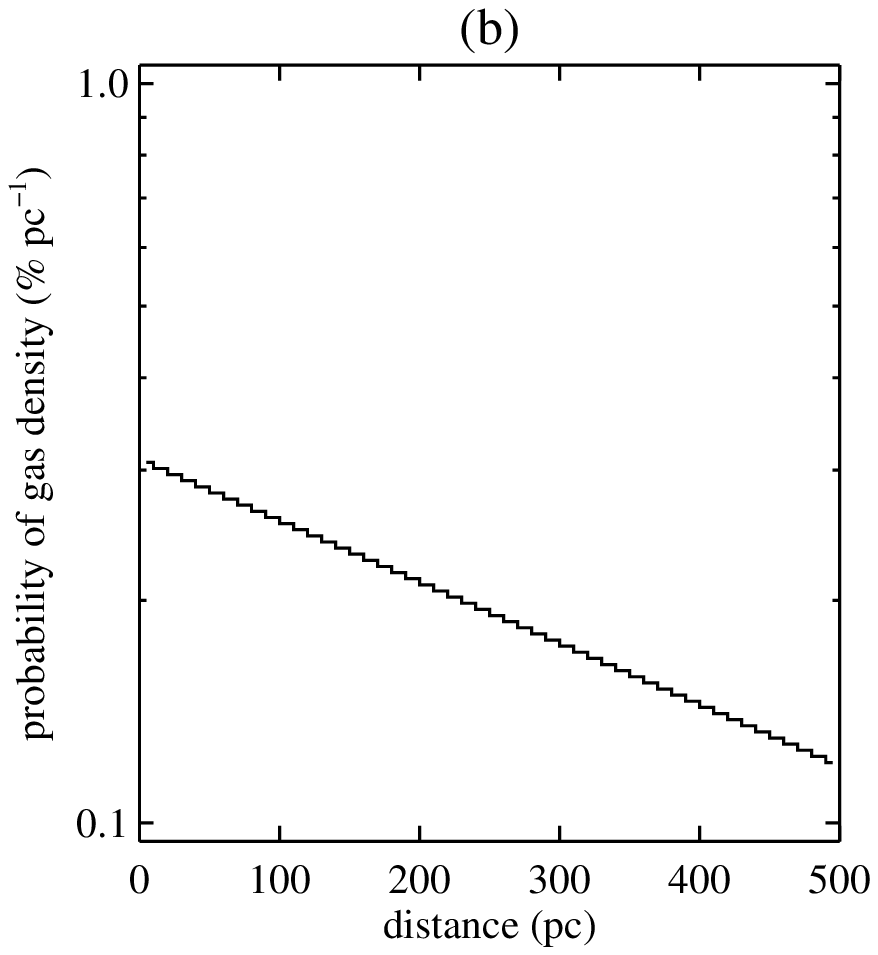}
	\end{center}
	\caption{
		Radial profiles of (a) the intensity of the radiation field, $I_{UV}$, and (b) gas density along the line of sight with Galactic coordinates (\textit{l}, \textit{b}) $\sim$ (0$\degr$, 15$\degr$). \label{fig:iuvsample}}
\end{figure*}

\begin{figure*}
	\begin{center}
		\includegraphics[trim=430 730 430 730,clip,height=15cm,angle=270]{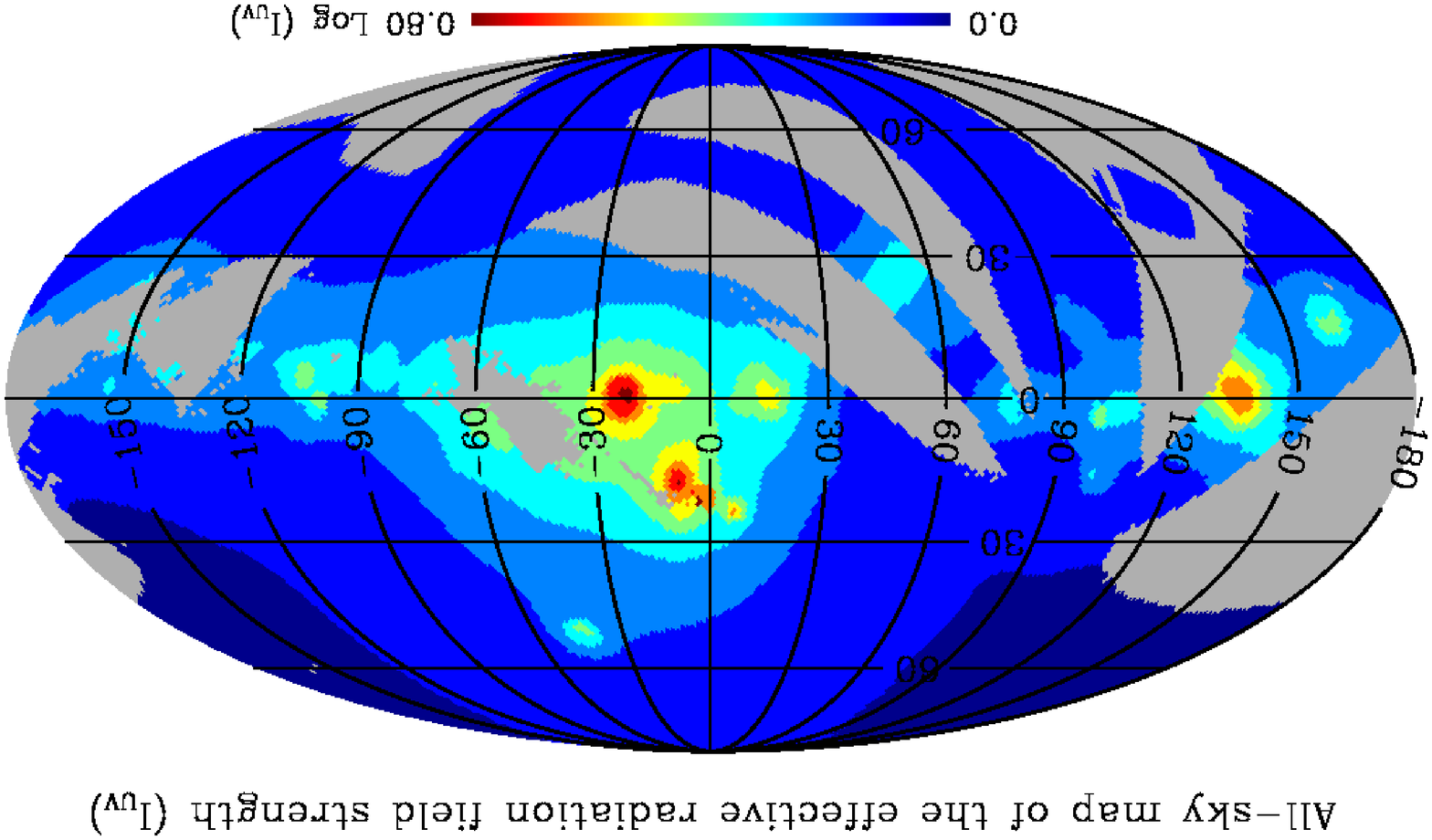}
	\end{center}
	\caption{
		All-sky map of the effective radiation field strength $I_{UV}$, obtained by weighting the 3D radiation field with the gas density profile and smoothing the map using the smoothing radius map in Figure \ref{fig:h2map}(d). \label{fig:iuv}}
\end{figure*}

\begin{figure*}
	\begin{center}
		\includegraphics[height=7.5cm]{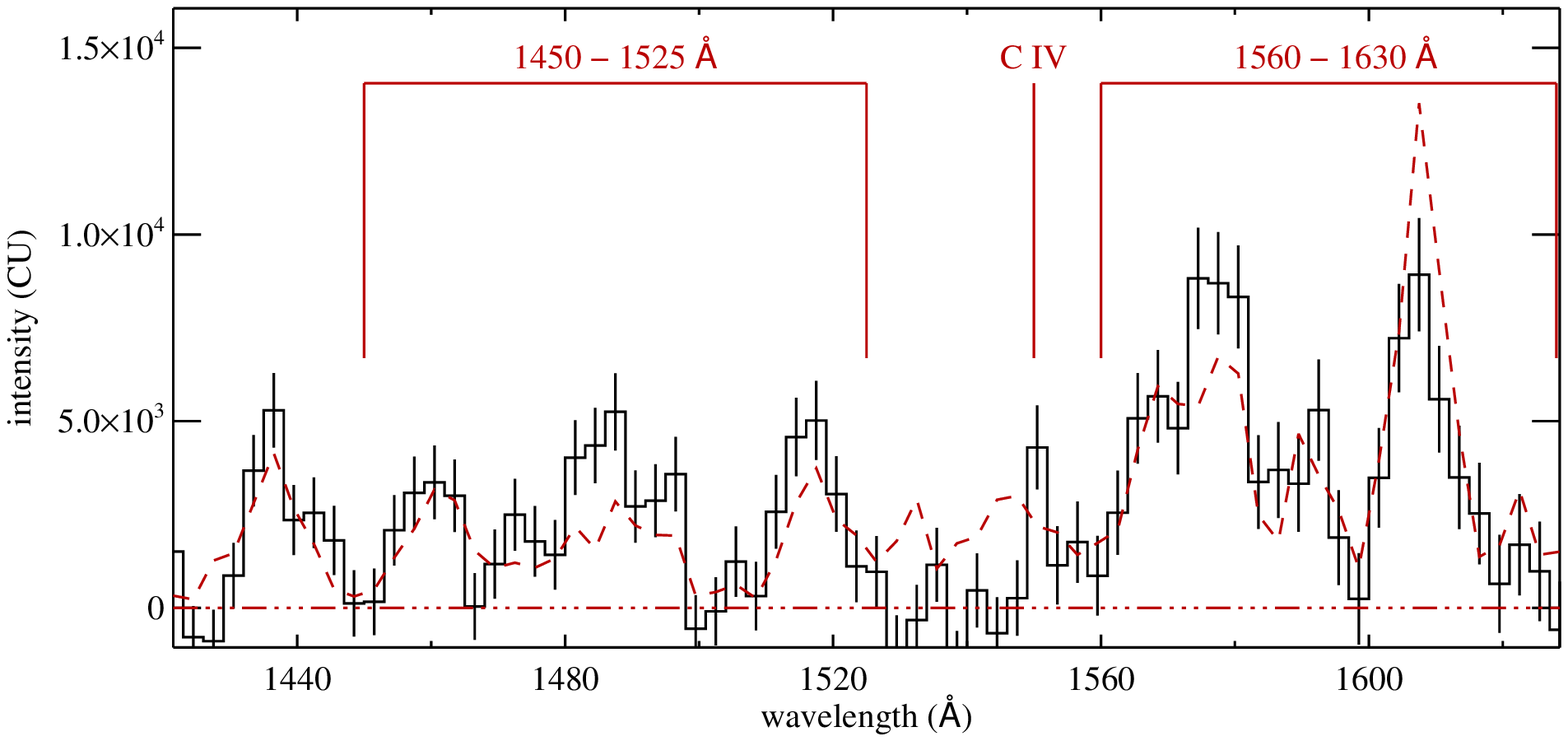}
	\end{center}
	\caption{
		Black solid line is the extinction-corrected spectrum of the FUV fluorescent H$_2$ emission lines at (\textit{l}, \textit{b}) $\sim$ (0$\degr$, 15$\degr$). Red dashed line is the best-fit PDR model spectrum with \textit{N}(H$_2$) = 10$^{20.1}$ cm$^{-2}$, $I_{UV}$=10$^{0.4}$, and a reduced $\chi^2$ value of 1.56. Horizontal red dash-dotted line at the bottom of the figure represents the continuum level. \label{fig:pdrspec}}
\end{figure*}

The variation of $I_{UV}$ as a function of distance at various sight lines was investigated by calculating $I_{UV}$ up to 500 pc from the Sun; $I_{UV}$ decreases sharply at distances greater than 500 pc. Figure \ref{fig:iuvsample}(a) shows the variation and three representative values--(1) the median (1.77), (2) the average (2.25), and (3) the weighted average (2.39)--of $I_{UV}$ at the line of sight with Galactic coordinates (\textit{l}, \textit{b}) $\sim$ (0$\degr$, 15$\degr$). The two peaks observed at about 100 pc and 200 pc are the result of contributions from the bright local stars in this particular direction: the peak at about 100 pc is the combined effect of Zeta Ophiuchi (HD149757) and V1067 Sco (HD151831), and the peak at about 200 pc is mainly due to Sigma Sco (HD147165). The weighted average of $I_{UV}$ was obtained by weighting the 3D mean intensity, $I_{UV}$, with the gas density profile, shown in Figure \ref{fig:iuvsample}(b), which was calculated by assuming a plane-parallel ISM. If the scale height of the plane-parallel ISM is 125 pc [i.e., the scale height of the dust layer in \citet{mar06}], then the gas density is given by $n(d)=n_0 e^{-d|\sin{b}|/125}$, where \textit{d} is the distance (in pc) from the Sun, \textit{b} is the Galactic latitude, and $n_0$ is the gas density at the Galactic plane (\textit{d} = 0). The gas density profile in Figure \ref{fig:iuvsample}(b) was normalized by integrating the gas density profile over distances ranging from 0 to 500 pc. Because both OB stars and the ISM are concentrated at the Galactic plane [i.e., the scale height of OB stars is 45 $\pm$ 20 pc \citep{ree00}], the H$_2$ fluorescence emission should originate mainly from the vicinity of the Galactic plane. Therefore, of the three representative values, the strength of the gas density-weighted radiation is the most reasonable value of $I_{UV}$. The average and median values tend to underestimate the strength of the incident UV radiation because of the rapid decrease of $I_{UV}$ with distance. Therefore, we adopted the gas density-weighted $I_{UV}$ as the effective $I_{UV}$. After creating the all-sky map of the effective $I_{UV}$ using $N_{side}$ = 64, we smoothed the map with the smoothing radius map in Figure \ref{fig:h2map}(d) to match the spatial resolution of the H$_2$ fluorescent emission map in Figure \ref{fig:h2map}(a). Finally, the effective $I_{UV}$ values were quantized in steps of 0.1 dex; the resulting map is shown in Figure \ref{fig:iuv}. The effective $I_{UV}$ in the Milky Way ranges from 0.0 to 0.8 dex. The effective $I_{UV}$ map in Figure \ref{fig:iuv} was used as input for the PDR model for each line of sight of the FUV fluorescent H$_2$ emission map.

\begin{figure*}
	\begin{center}
		\includegraphics[trim=430 730 430 730,clip,height=15cm,angle=270]{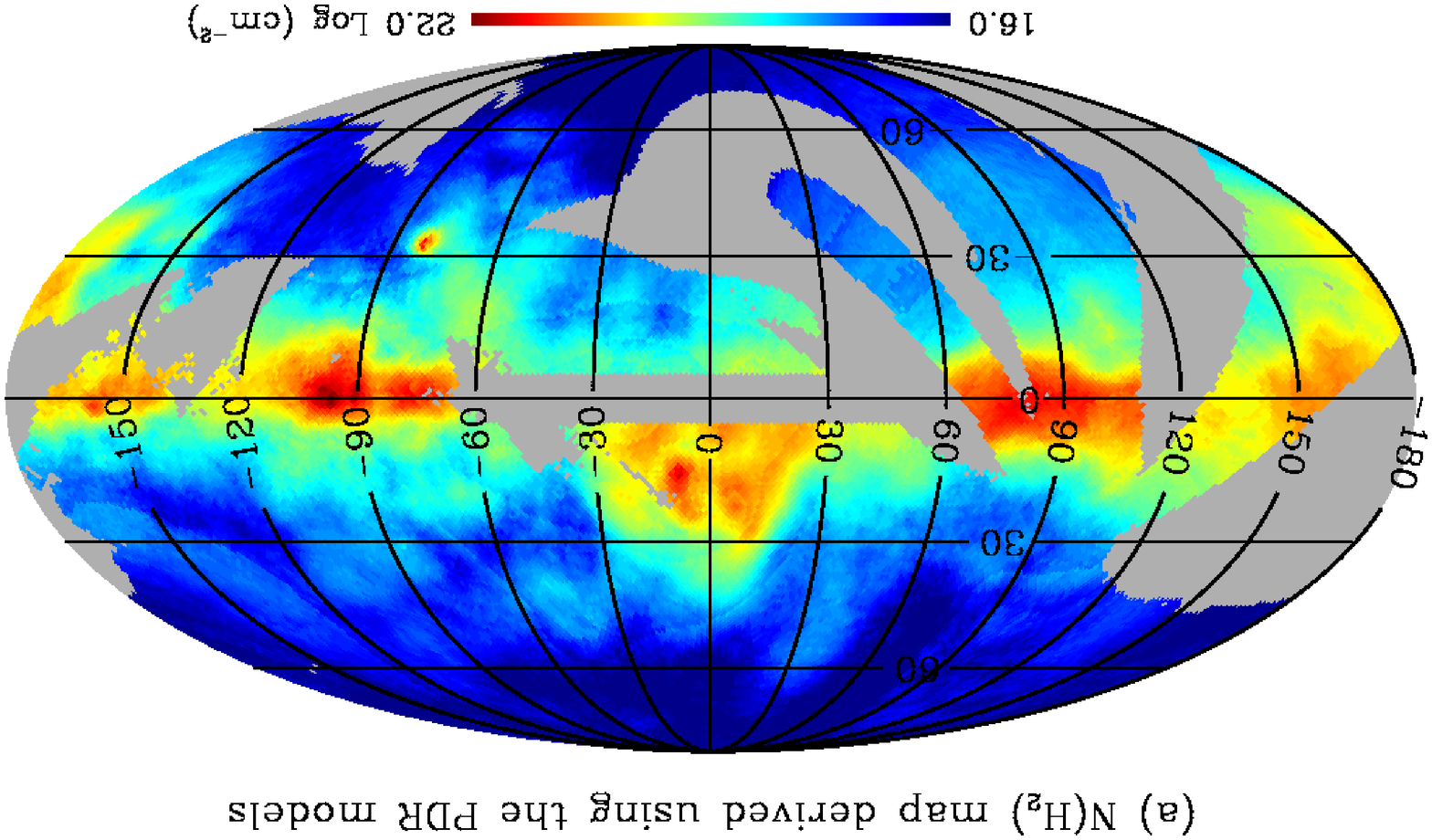}\\
		\vspace{10pt}
		\includegraphics[trim=430 730 430 730,clip,height=15cm,angle=270]{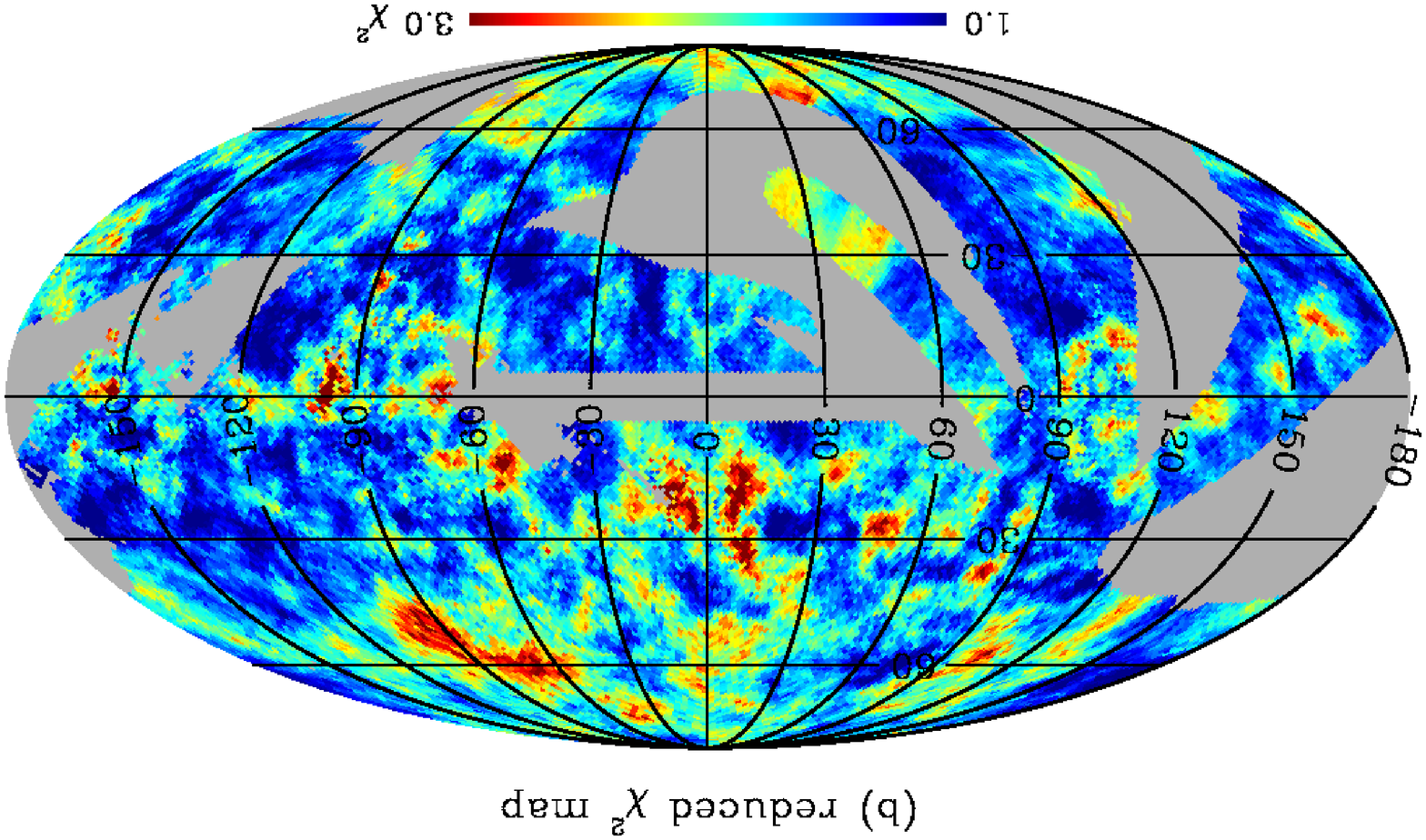}
	\end{center}
	\caption{
		All-sky maps of (a) \textit{N}(H$_2$) derived using the PDR models and (b) the reduced $\chi^2$ value. \label{fig:nh2map}}
\end{figure*}

\begin{figure*}
	\begin{center}
		\includegraphics[height=7.5cm]{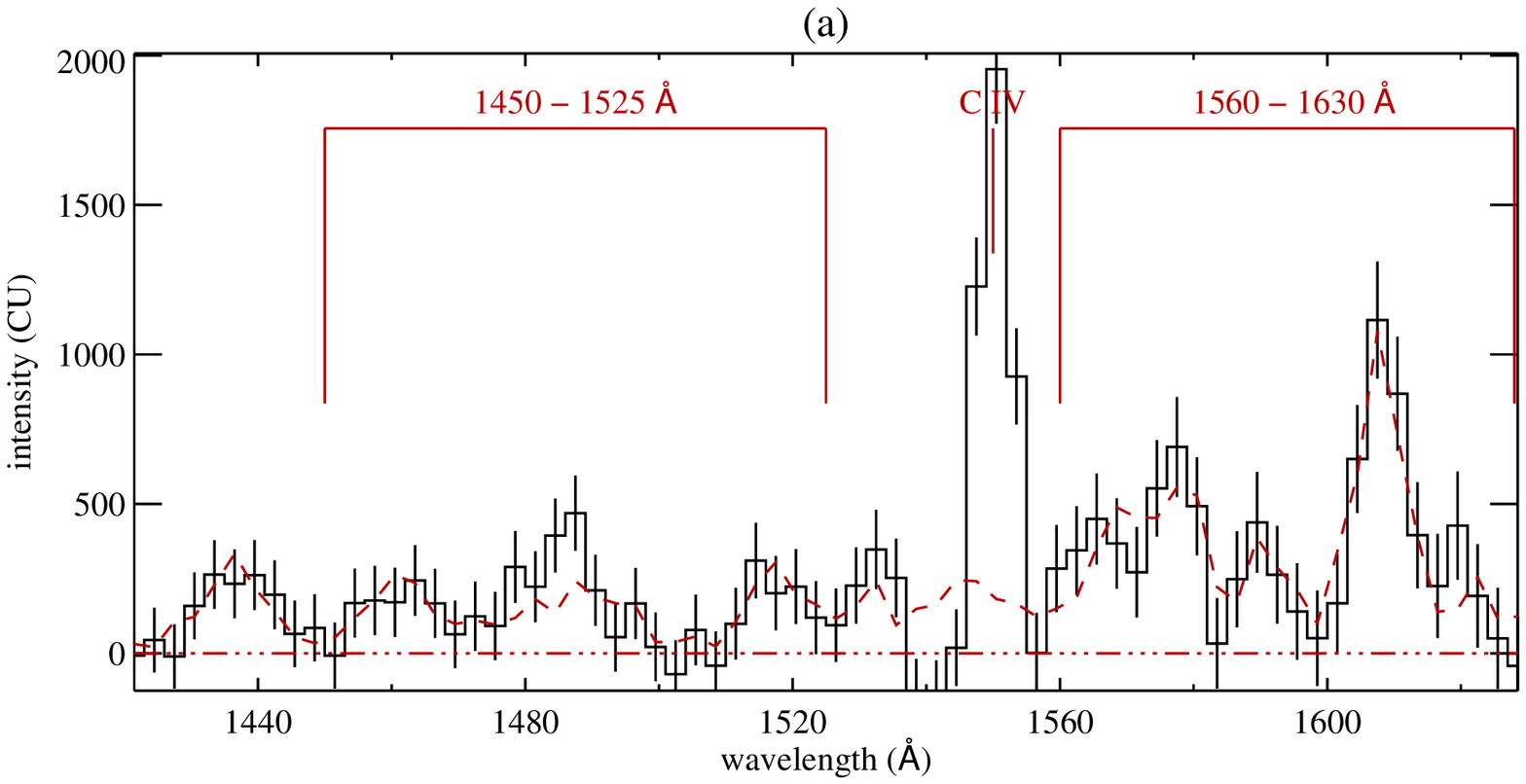}\\
		\includegraphics[height=7.5cm]{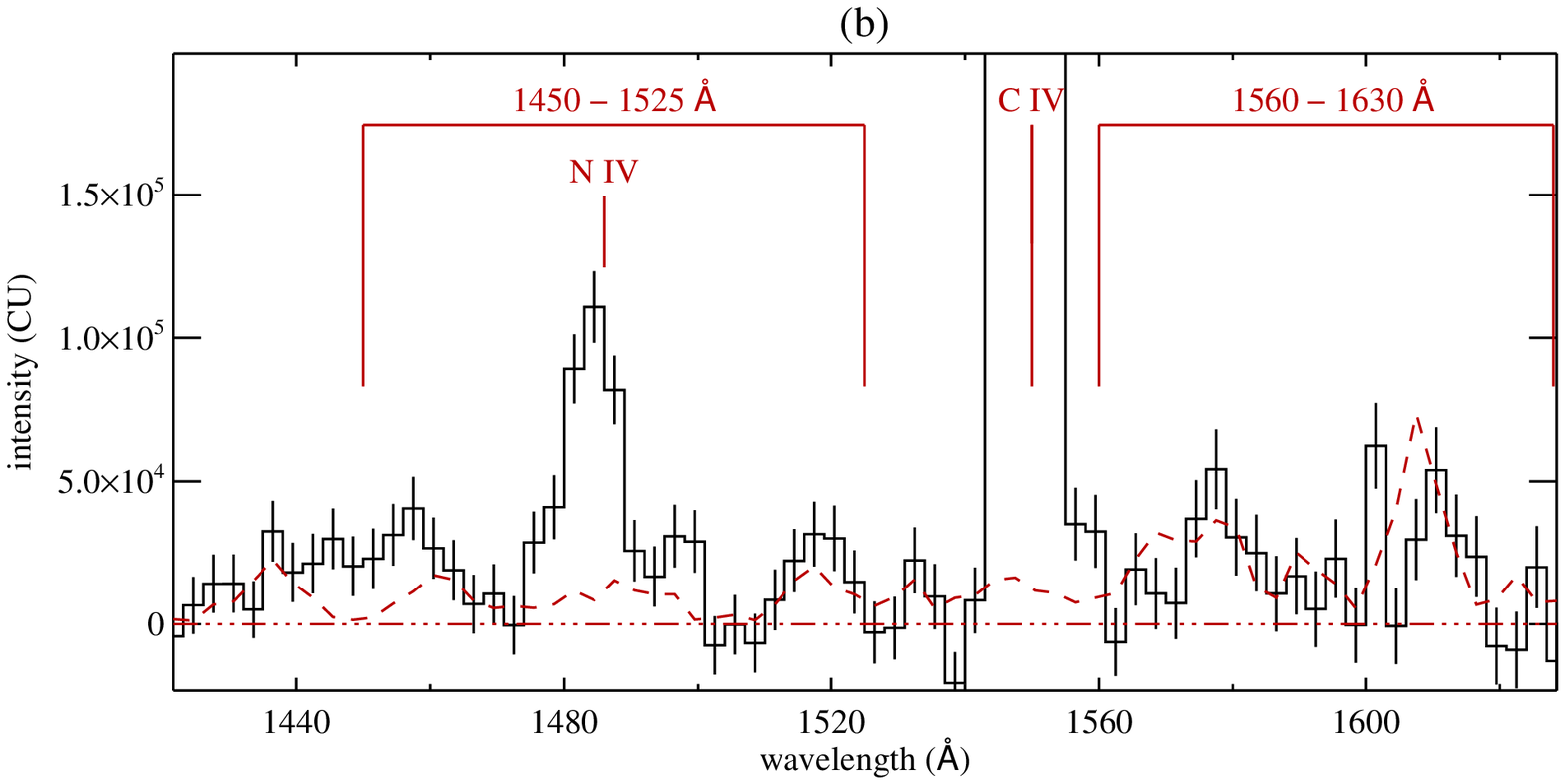}		
	\end{center}
	\caption{
		Two examples of the PDR fitting result: (a) For (\textit{l}, \textit{b}) $\sim$ (19$\degr$, 29$\degr$) near the Ophiuchus cloud (reduced $\chi^2$ = 0.57) and (b) for (\textit{l}, \textit{b}) $\sim$ (262$\degr$, 0$\degr$) near the Vela supernova remnant (reduced $\chi^2$ = 4.96). Black solid lines represent the extinction-corrected spectra of the FUV fluorescent H$_2$ emission lines, and red dashed lines represent the best-fit PDR model spectra. \label{fig:pdrchi}}
\end{figure*}

The H$_2$ fluorescence spectra were generated by varying log \textit{N}(H$_2$) from 13.5 to 25.0 in steps of 0.1 for the relevant effective $I_{UV}$ value for each line of sight shown in Figure \ref{fig:iuv}. We used a $\chi^2$ minimization method to find the best-fit \textit{N}(H$_2$) for each sight line. Figure \ref{fig:pdrspec} shows an example of the model fit for the line of sight in Figures \ref{fig:samplespec} and \ref{fig:iuvsample}. The black solid line in Figure \ref{fig:pdrspec} is the extinction-corrected emission line spectrum, whereas the blue solid line in Figure \ref{fig:samplespec} is the observed emission line spectrum before correcting for dust extinction. The all-sky \textit{N}(H$_2$) map constructed from the PDR model is shown in Figure \ref{fig:nh2map}(a) and its reduced $\chi^2$ map is shown in Figure \ref{fig:nh2map}(b). The reduced $\chi^2$ value ranges from 0.57 to 4.96 with an average of 1.65 $\pm$ 0.40. In general, the fitting was quite good for the molecular clouds, whereas it was quite poor in the regions of high-temperature gas. Figure \ref{fig:pdrchi} displays two examples of the fitting result. Figure \ref{fig:pdrchi}(a), the best case with the reduced $\chi^2$ = 0.57 obtained for the location of (\textit{l}, \textit{b}) $\sim$ (19$\degr$, 29$\degr$) near the Ophiuchus cloud, shows excellent fitting for the molecular clouds. Figure \ref{fig:pdrchi}(b) is the worst case with the reduced $\chi^2$ = 4.96 obtained for the location of (\textit{l}, \textit{b}) $\sim$ (262$\degr$, 0$\degr$) near the Vela supernova remnant where the H$_2$ fluorescence emission is much weaker than the atomic line emissions. For this particular case, the \mbox{N\,{\sc iv}} line, whose intensity is generally below the level of nearby H$_2$ fluorescence emission lines in other regions of the sky and thus not excluded in the PDR fitting, greatly increases the $\chi^2$ value. Nevertheless, the fitting result of the H$_2$ fluorescence emission lines seems quite good even for this extreme case.

\begin{figure*}
	\begin{center}
		\includegraphics[trim=430 730 430 730,clip,height=15cm,angle=270]{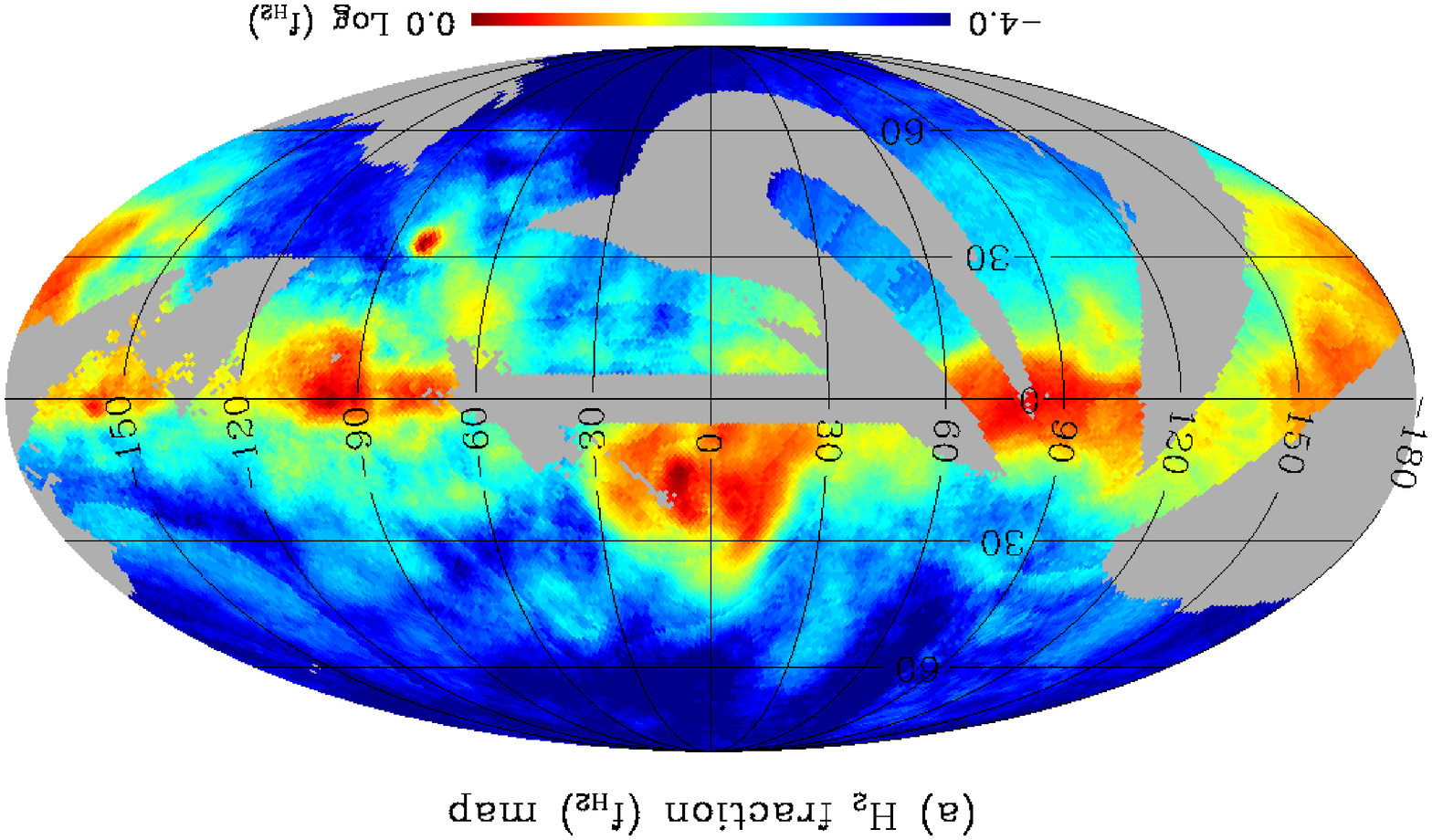}\\
		\vspace{10pt}
		\includegraphics[height=7.5cm]{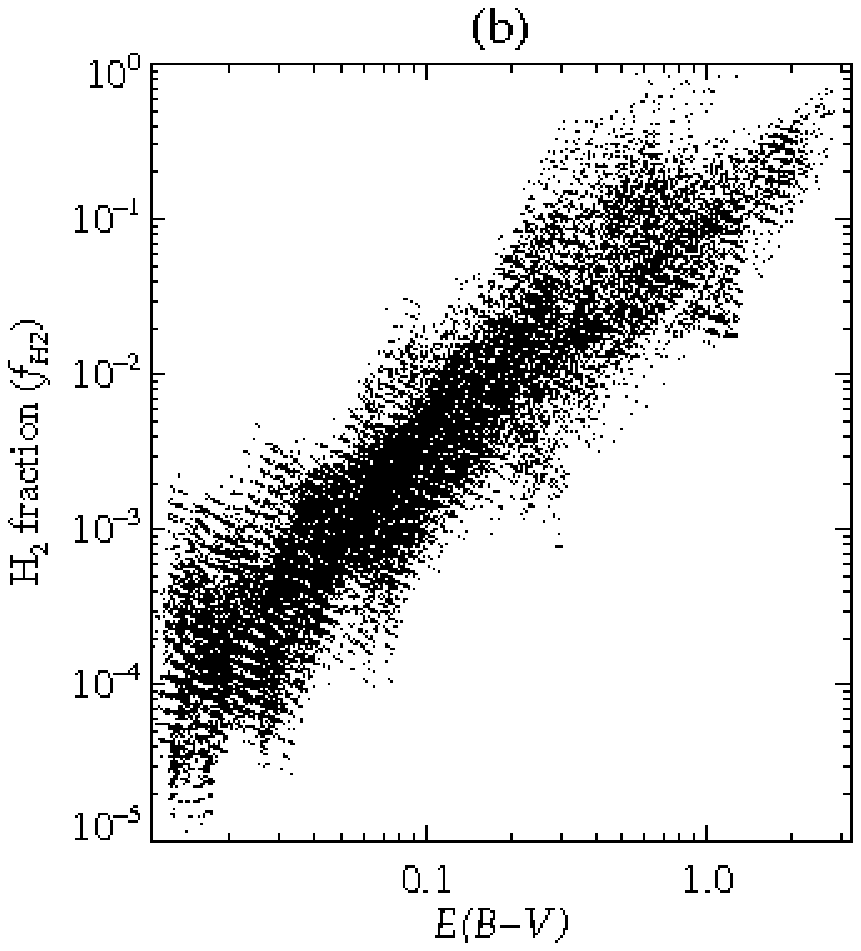}\hspace{10pt}
		\includegraphics[height=7.5cm]{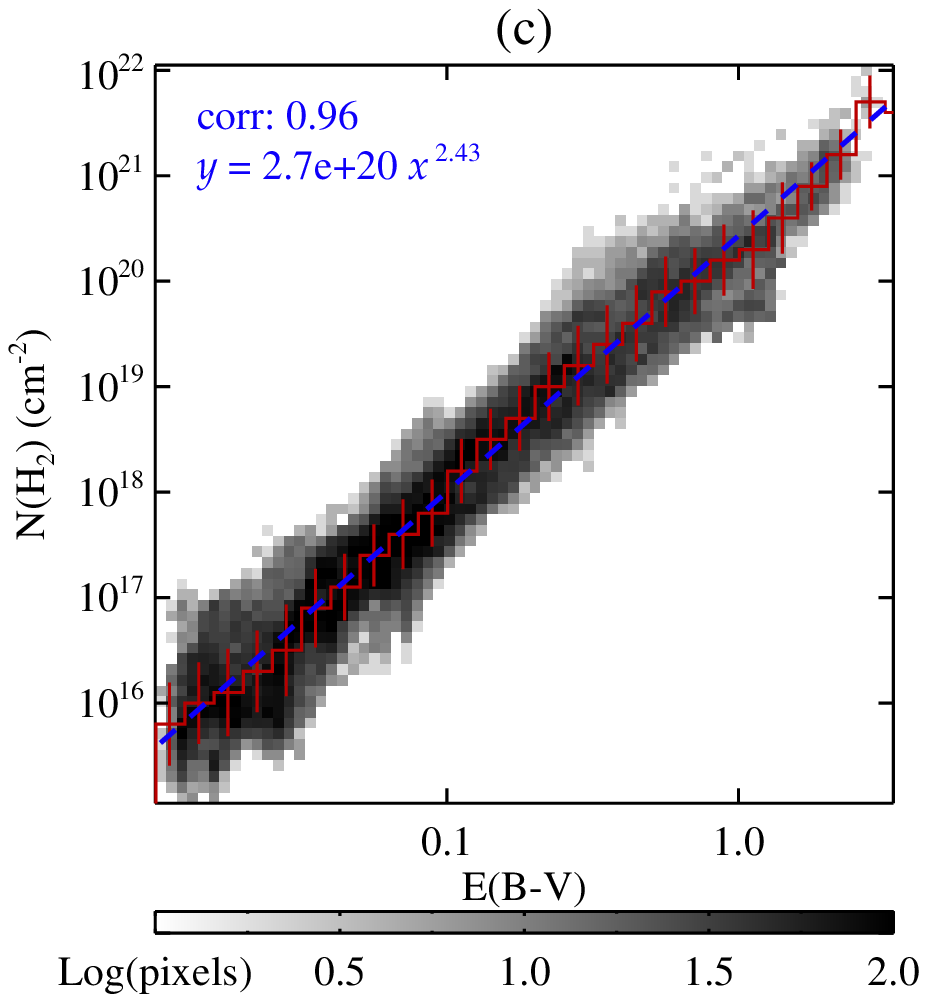}
	\end{center}
	\caption{
		(a) All-sky map of the H$_2$ fraction $f_{H2}$. (b) Scatterplot of $f_{H2}$ as a function of \textit{E(B\textendash V)}. (c) Correlation between \textit{N}(H$_2$) and \textit{E(B\textendash V)} using a colorbar representing the number of pixels. \label{fig:h2fraction}}
\end{figure*}

The all-sky \textit{N}(H$_2$) map in Figure \ref{fig:nh2map}(a) is very similar to the FUV fluorescent H$_2$ emission map in Figure \ref{fig:extcorr}(a). However, the dynamic range of \textit{N}(H$_2$) in Figure \ref{fig:nh2map}(a) is remarkably different from that of the FUV fluorescent H$_2$ emission in Figure \ref{fig:extcorr}(a). The \textit{N}(H$_2$) ranges from 10$^{16}$ to 10$^{23}$ cm$^{-2}$ while the H$_2$ intensity ranges from 10$^{3.5}$ to 10$^{6.4}$ CU. The wide dynamic range of \textit{N}(H$_2$) is related to the wide dynamic range of H$_2$ fraction in the Milky Way, defined as $f_{H2}$ = 2\textit{N}(H$_2$)/[2\textit{N}(H$_2$) + \textit{N}(\mbox{H\,{\sc i}}))] and shown in Figure \ref{fig:h2fraction}(a) and (b). At optically thin regions where \textit{E(B\textendash V)} $<$ 0.1, $f_{H2}$ $<$ 1$\%$. However, $f_{H2}$ gradually increases as \textit{E(B\textendash V)} increases, i.e., $f_{H2}$ is $\sim$10$\%$ when \textit{E(B\textendash V)} = 1 and $\sim$50$\%$ when \textit{E(B\textendash V)} = 3 in the Galactic plane. At optically thick regions, UV radiation is strongly attenuated, thus shielding the H$_2$ molecules from the photodissociation effect and allowing $f_{H2}$ to increase with increasing \textit{E(B\textendash V)}. Moreover, the H$_2$ formation rate via dust grain catalysis is more active in denser dust clouds. Figure \ref{fig:h2fraction}(c) presents the correlation between \textit{N}(H$_2$) and \textit{E(B\textendash V)}. The red solid lines represent the median and standard deviation of \textit{N}(H$_2$) for a step size of 0.1 dex of the \textit{E(B\textendash V)} interval, and the blue dashed line indicates the best-fit power law function. Note that the correlation between \textit{N}(H$_2$) and \textit{E(B\textendash V)} is strong with a correlation factor of 0.96. The standard deviation of \textit{N}(H$_2$) from the fitting line is about $\pm$ 0.5 dex.

\begin{figure*}
	\begin{center}
		\includegraphics[trim=430 730 430 730,clip,height=15cm,angle=270]{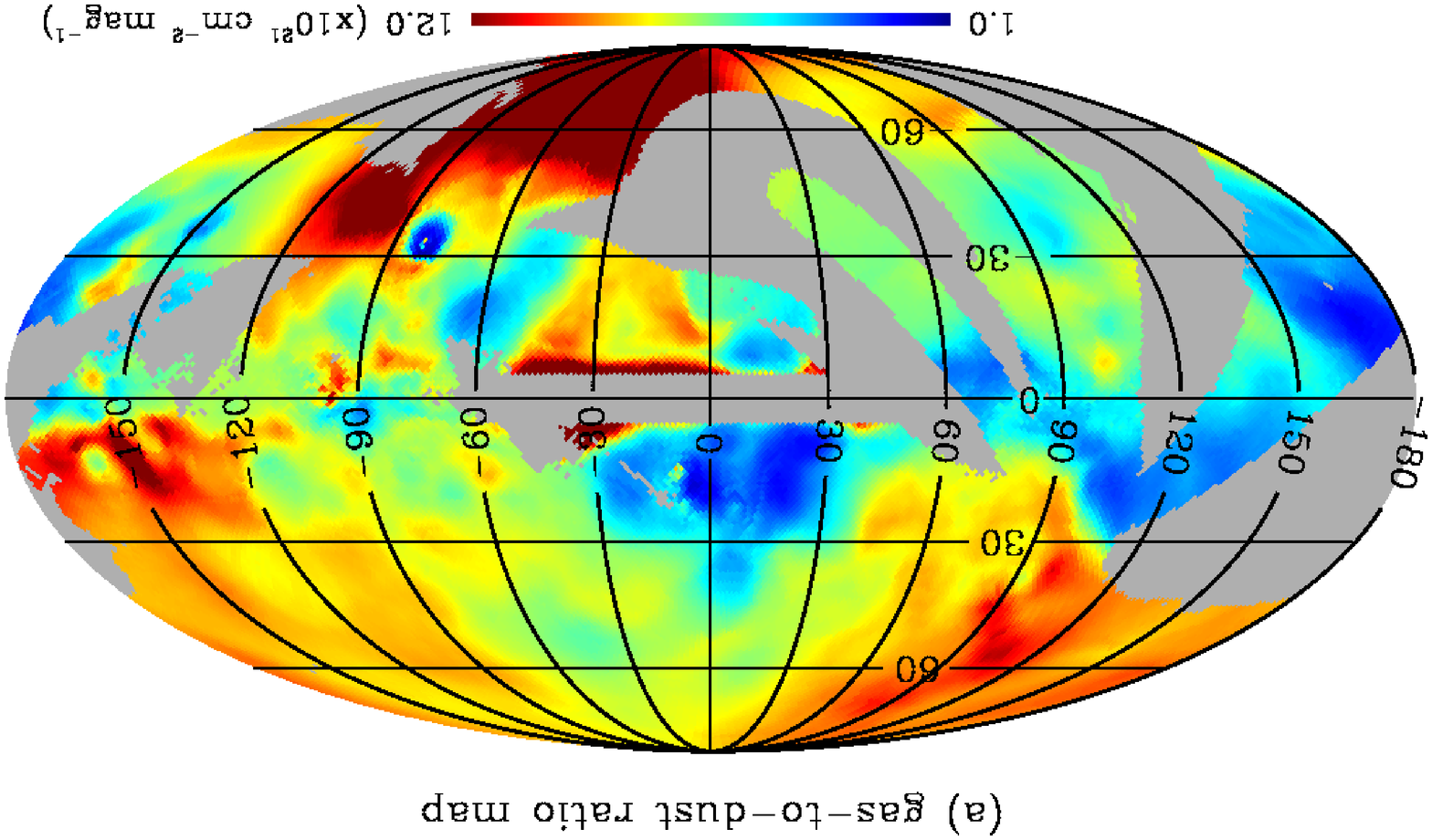}\\
		\vspace{10pt}
		\includegraphics[height=7.5cm]{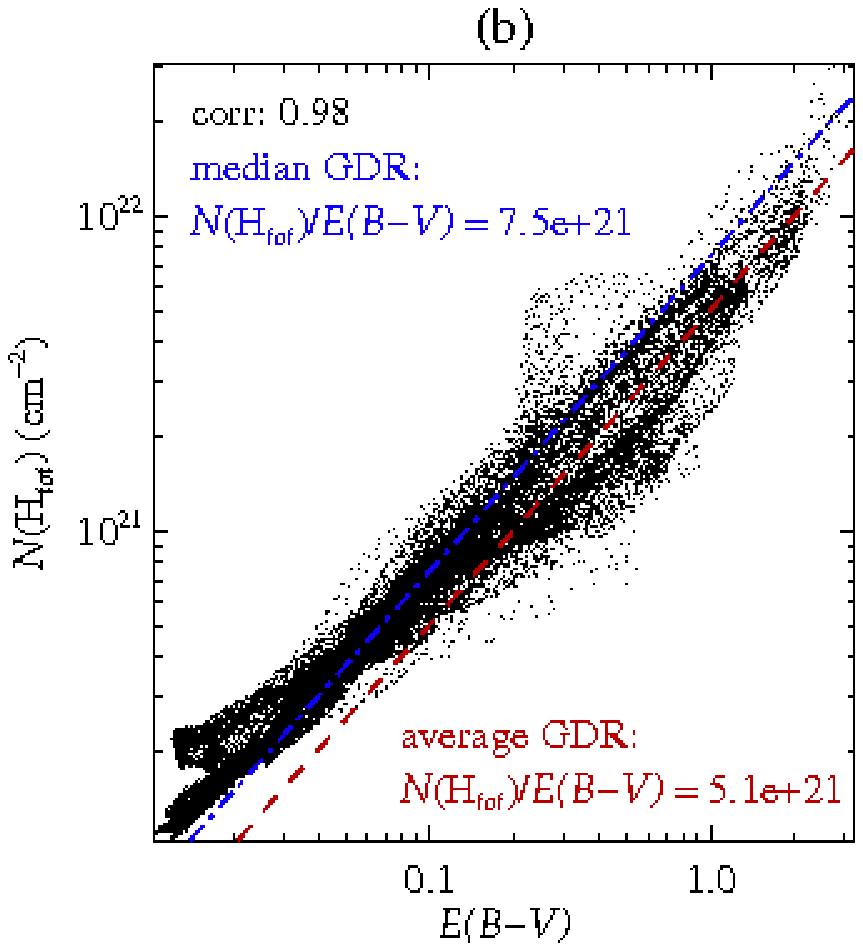}
	\end{center}
	\caption{
		(a) All-sky map of GDR. (b) Scatterplot of \textit{N}(H$_{tot}$) as a function of \textit{E(B\textendash V)}. \label{fig:gdr}}
\end{figure*}

\subsection{Calculation of GDR}

\citet{boh78} used the \textit{Copernicus} observations of 75 stars to obtain the mean ratio of the total amount of neutral hydrogen to \textit{E(B\textendash V)}, $\langle$\textit{N}(\mbox{H\,{\sc i}} + H$_2$)/\textit{E(B\textendash V)}$\rangle$ = 5.8 $\times$ 10$^{21}$ atoms cm$^{-2}$ mag$^{-1}$. \citet{rac09} found that the GDR calculated using the Far Ultraviolet Spectroscopic Explorer (\textit{FUSE}) observations of 38 translucent lines of sight is essentially identical to the \textit{Copernicus} value.

While these previous studies were based on the absorption lines of bright stars, we took a different approach of using the FUV H$_2$ fluorescence emission combined with the PDR model to estimate the GDR. The total hydrogen column density \textit{N}(H$_{tot}$) was calculated by adding \textit{N}(\mbox{H\,{\sc i}}) in Figure \ref{fig:othermap}(b) and 2\textit{N}(H$_2$) in Figure \ref{fig:nh2map}(a); it is plotted as a function of \textit{E(B\textendash V)} in Figure \ref{fig:gdr}(b). The all-sky GDR map calculated using our results is shown in Figure \ref{fig:gdr}(a). The red dashed line in Figure \ref{fig:gdr}(b) is the best fit of a linear function with zero intercept, and the slope (5.1 $\times$ 10$^{21}$ atoms cm$^{-2}$ mag$^{-1}$) of the linear function is regarded as an average GDR. The blue dash-dotted line in Figure \ref{fig:gdr}(b) indicates the median GDR value in the all-sky GDR map [Figure \ref{fig:gdr}(a)], which is (7.5 $\pm$ 2.4) $\times$ 10$^{21}$ atoms cm$^{-2}$ mag$^{-1}$. These differently derived GDRs are consistent with the standard value and are within the range of uncertainty despite the many assumptions made for the estimation of \textit{N}(H$_2$) from its fluorescence emission such as the averaged interstellar radiation field and the single-cloud model. The scatter seen in Figures \ref{fig:gdr}(a) and (b) may be partly due to the intrinsic variation of the GDR and to the assumptions made in correcting the dust extinction for the FUV spectra, those made in the PDR modeling, or both. We note in Figure \ref{fig:gdr}(b) that the data points reside above the slopes in the region of \textit{E(B\textendash V)} well below 0.1, which corresponds to the halo region, implying that the gas component is more abundant in the halo region than the estimated GDR predicts.


\section{Summary}

We presented the first all-sky map of FUV fluorescent H$_2$ emission of the Milky Way Galaxy, covering approximately 76$\%$ of the sky. On average, the H$_2$ fluorescence emission contributed $\sim$8.7 $\pm$ 2.4$\%$ of the FUV flux in the wavelength range of 1370--1710 {\AA} observed by \textit{FIMS/SPEAR}, except in the regions near the Ophiuchus cloud and the Eridanus region, where the ratio of the H$_2$ intensity to the total FUV intensity is much higher. After correcting for dust extinction, the H$_2$ fluorescence emission clearly correlates with \textit{E(B\textendash V)}, \textit{N}(\mbox{H\,{\sc i}}), and H$\alpha$ emission. The spatial distribution of H$_2$ was obtained from the H$_2$ fluorescence emission map using a simple plane-parallel PDR model with a uniform pressure of \textit{p}/\textit{k} = 1000 cm$^{-3}$ K and the interstellar radiation field estimated from the luminosities of the UV stars. According to the model, \textit{N}(H$_2$) ranges from 10$^{16}$ to 10$^{23}$ cm$^{-2}$. The fraction of H$_2$ was found to be $<$1$\%$ at optically thin regions when \textit{E(B\textendash V)} $<$ 0.1 but gradually increases to $\sim$50$\%$ when \textit{E(B\textendash V)} = 3 in the Galactic plane, implying strong self-shielding by H$_2$ and active H$_2$ formation there. In addition, the GDR was found to be $\sim$5.1 $\times$ 10$^{21}$ atoms cm$^{-2}$ mag$^{-1}$, which is consistent with the standard value of 5.8 $\times$ 10$^{21}$ atoms cm$^{-2}$ mag$^{-1}$ derived from the \textit{Copernicus} and \textit{FUSE} observations and is within the uncertainty range. However, we found that the GDR is higher than the standard value in the Galactic halo region.

\acknowledgments

\textit{FIMS/SPEAR} was funded by the Ministry of Science and Technology (Korea) and a grant NAG5-5355 (NASA). It is a joint project of the Korea Astronomy and Space Science Institute, the Korea Advanced Institute of Science and Technology, and the University of California at Berkeley (USA). We acknowledge the use of the Legacy Archive for Microwave Background Data Analysis (LAMBDA), part of the High Energy Astrophysics Science Archive Center (HEASARC). HEASARC/LAMBDA is a service of the Astrophysics Science Division at the NASA Goddard Space Flight Center. This research was supported by the Korea Astronomy and Space Science Institute under the R$\&$D program supervised by the Ministry of Science, ICT and Future Planning of Korea. This research was also supported by BK 21 plus program and Basic Science Research Program (2017R1D1A1B03031842) through the National Research  foundation (NRF) funded by the Ministry of Education of Korea.

\clearpage

\end{document}